\documentclass[numbib]{article_class}%%%%where imaiai is the template name
\usepackage{amsthm,amsmath}
\usepackage[utf8]{inputenc} %unicode support
\usepackage{bm}% bold math
\usepackage{amsfonts}
\usepackage{amssymb}
\usepackage{graphicx}% Include figure files
\usepackage{color}

\usepackage{array}
\newcolumntype{L}[1]{>{\raggedright\let\newline\\\arraybackslash\hspace{0pt}}m{#1}}
\newcolumntype{C}[1]{>{\centering\let\newline\\\arraybackslash\hspace{0pt}}m{#1}}
\newcolumntype{R}[1]{>{\raggedleft\let\newline\\\arraybackslash\hspace{0pt}}m{#1}}

\begin{document}

\title{Influencer identification in dynamical complex systems}

%\shorttitle{Influencer identification} %%%for recto running head
%\shortauthorlist{S. Pei et al.} %%% for verso running head

\author{{%%%% First author details
\sc Sen Pei}$^{*,\dagger}$,\\[2pt]
Department of Environmental Health Sciences,\\
Mailman School of Public Health, Columbia University,\\
New York, NY 10032, USA\\[2pt]
%%%%%%% 
{\sc Jiannan Wang}$^\dagger$\\[2pt]
Research Institute of Frontier Science, \\
Beihang University, Beijing 100191, China\\[2pt]
Levich Institute and Physics Department, \\
City College of New York, New York, NY 10031, USA\\[2pt]
%%%%%%%
{\sc Flaviano Morone}\\[2pt]
Levich Institute and Physics Department, \\
City College of New York, New York, NY 10031, USA\\[6pt]
%%%%%%%
{\sc and}\\[6pt]
{\sc Hern\'{a}n A Makse}$^*$ \\[2pt]
Levich Institute and Physics Department, \\
City College of New York, New York, NY 10031, USA\\[6pt]
$^*${sp3449@cumc.columbia.edu (SP), hmakse@lev.ccny.cuny.edu (HAM)}\\[2pt]
$^\dagger${These authors contributed equally to this work.}
}

\maketitle

\begin{abstract}
{The integrity and functionality of many real-world complex systems hinge on a small set of pivotal nodes, or influencers. In different contexts, these influencers are defined as either structurally important nodes that maintain the connectivity of networks, or dynamically crucial units that can disproportionately impact certain dynamical processes. In practice, identification of the optimal set of influencers in a given system has profound implications in a variety of disciplines. In this review, we survey recent advances in the study of influencer identification developed from different perspectives, and present state-of-the-art solutions designed for different objectives. In particular, we first discuss the problem of finding the minimal number of nodes whose removal would breakdown the network (i.e., the optimal percolation or network dismantle problem), and then survey methods to locate the essential nodes that are capable of shaping global dynamics with either continuous (e.g., independent cascading models) or discontinuous phase transitions (e.g., threshold models). We conclude the review with a summary and an outlook.}
\end{abstract}

\section{Introduction}

A wide variety of phenomena in nature and society can be unified under the umbrella of dynamical complex systems. Important social and biological processes such as epidemic outbreaks in population \cite{pastor2015epidemic}, information diffusion in social media \cite{zhang2016dynamics}, signal transmission in brain networks \cite{bullmore2009complex} and dynamical evolution of ecosystems \cite{montoya2006ecological} all boil down to interactions among large numbers of building units of each system, and therefore can be properly described by dynamical models in complex networks \cite{newman2003structure,barrat2008dynamical,boccaletti2006complex,albert2002statistical}. In these systems, complex interactions at microscopic scale lead to the abundant dynamical behaviors we observe at macroscopic level. As a result, understanding how network structure impacts the function of dynamical complex systems becomes a central topic in modern network science. 

\begin{figure}[]
 \centering
  \includegraphics[width=0.9\columnwidth]{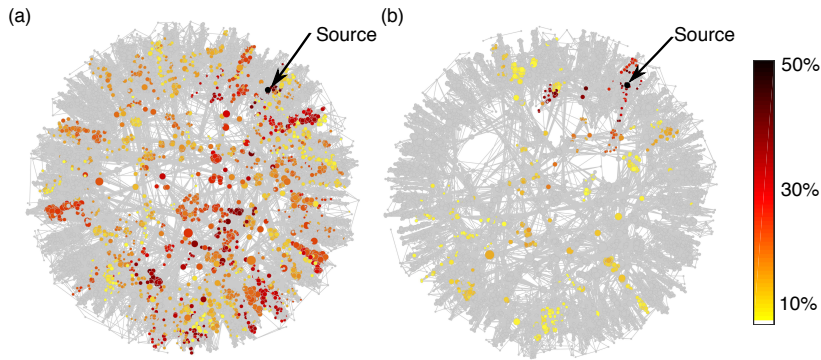}
  \caption{{Simulations of epidemic processes initiated from different origins.} We use a susceptible-infected-removed (SIR) model to simulate epidemic outbreaks in a patient-to-patient contact network from multiple Swedish hospitals \cite{Pei2019inference}. Given the same set of epidemiological parameters (infectious rate $\beta=0.01$ and recover rate $\mu=1$), outbreaks initiated from two origins have dramatically different outcomes. In (a) and (b), the epidemic sources have the same number of connections, but the source in (a) locates in a more central region with a higher k-core index. The color of each node indicates the probability of infection during 1,000 independent realizations of the SIR model. } \label{SIRsim}
      \end{figure}

In network science, it has been well established that the collective dynamics of complex systems can be shaped by a small number of essential nodes, or influencers. For example, opinion leaders in social media are capable of influencing the public viewpoint on certain trending topics \cite{watts2007influentials}; critical regions in brain are essential in the formation of memory networks \cite{bullmore2009complex,del2018finding,reis2014avoiding,zamora2010cortical}; and keystone species in ecology are responsible for the integrity and stability of ecosystems \cite{may1972will,scheffer2012anticipating,mills1993keystone,morone2018kcore}. Numerical simulations of epidemic processes have also demonstrated that the location of epidemic origin is critical for the final outbreak size \cite{kitsak2010identification} (see Figure \ref{SIRsim} for an example). 

In general, influencers can be vaguely defined as the nodes that are disproportionately ``important'' to the function of complex systems. However, in a given context, influencers may have a more specific definition: in social networks, influencers are opinion leaders who can influence a large number of people; in brain networks, influencers are important regions that maintain the connection across different functional parts; in ecological systems, influencers are keystone species whose extinction would collapse the network; and in epidemic spreading, influencers are superspreaders who transmit infectious diseases to a large population. In previous studies, abundant works exist dedicated to explore how to find influencers in a specific system. (For instance, in social science, various centrality measures have been developed to rank users' importance in social networks \cite{freeman1978centrality}.) Due to its vast scope, here we do not attempt to summarize all relevant works, but instead focus on two important problems with wide applications. 
\begin{itemize}
	\item First, how to find the {\it structural} influencers whose removal would fragment the network? This problem, named optimal percolation \cite{morone2015influence} or network dismantle \cite{braunstein2016network}, is purely structural and does not involve with dynamical processes.
	\item Second, how to find the {\it dynamical} influencers who can lead to the largest cascading following a spread model? This problem, named influence maximization \cite{kempe2003maximizing}, depends on both network structure and dynamical rules. 
\end{itemize}
In our following discussion, we refer to the above two problems uniformly as influencer identification. The specific definition of influencers is thus context-dependent. Solutions to these two problems can be applied in real-world applications ranging from maximization of marketing in social networks \cite{kempe2003maximizing,leskovec2007dynamics,richardson2002mining}, optimization of immunization campaigns \cite{pastor2002immunization,chen2008finding,cohen2003efficient} to protection of networks under malicious attacks \cite{albert2000error,cohen2001breakdown,latora2005vulnerability}. 

Real-world dynamical complex systems generally fall into two major classes:
\begin{itemize}
	\item {\it Systems with only positive interactions.} For instance, in online social media, social ties can facilitate the spread of information among users \cite{watts2007influentials}; in human population, physical contacts may transmit infectious diseases from person to person \cite{keeling2011modeling}; and in mutualistic ecosystems, cooperations between different species benefit their existence in ecology \cite{thebault2010stability}.
	\item {\it Systems with both positive and negative interactions.} For instance, in neural systems, synaptic connections can be either excitatory or inhibitory \cite{kandel2000principles}; in gene regulatory networks, molecular regulators can activate or inhibit the expression of certain genes \cite{davidson2005gene};  and in ecosystems, both mutualistic and predator-prey relationships coexist among different species \cite{thebault2010stability}. 
\end{itemize}
For systems with only positive interactions, influencer identification can be defined using key topological structures such as the giant component (GC) \cite{erdos1960evolution,callaway2000network,newman2002spread} and k-core \cite{seidman1983network,dorogovtsev2006k}. On the contrary, systems with both positive and negative interactions do not admit the classical definitions of the GC and k-core, so the influencer identification problem in these systems need to be treated with a different theory. In this review, we only consider the former case where all links have positive interactions, and the case of inhibition/activation interactions will be treated elsewhere.

For structural influencer identification, the solution only depends on the network structure. However, for dynamical influencer identification, spread models can be further divided into two classes with continuous (second order) and discontinuous (first order) phase transitions. In regular percolation process \cite{erdos1960evolution} and independent cascade models \cite{kermack1932contributions}, the GC emerges continuously from zero size as links are gradually occupied \cite{dorogovtsev2008critical}. In contrast, in k-core percolation \cite{baxter2010bootstrap} and threshold models \cite{granovetter1978threshold}, k-core structure with non-zero size can appear abruptly as more nodes are activated \cite{goltsev2006k,watts2002simple}. For these two types of dynamical models, approaches to find influencers are qualitatively different. We therefore discuss the influencer identification problem for models with continuous and discontinuous phase transitions separately.

Heuristically, influencers can be selected by picking vital individual spreaders one by one using a greedy approach, in which the influence of single nodes is estimated via Monte Carlo simulations \cite{kempe2003maximizing} or various centrality measures \cite{freeman1978centrality}. However, influencer identification is intrinsically an NP-hard combinatorial optimization problem \cite{kempe2003maximizing}. Therefore, a collective point of view that considers interactions among multiple spreaders is required. Recent progresses have translated the influencer identification problem into other closely related optimization problems such as message passing \cite{altarelli2014containing}, belief propagation \cite{altarelli2013optimizing}, optimal percolation \cite{morone2015influence}, optimal decycling \cite{braunstein2016network,mugisha2016identifying} and explosive percolation \cite{clusella2016immunization}. These new approaches have enriched our understanding of feasible directions to tackle the influencer identification problem, and provided a number of sophisticated yet efficient methods that are applicable to large-scale complex systems. Classical centrality-based approaches have been extensively discussed in previous literature. As a result, in this survey, we focus on development from other approaches. Readers who are interested in centralities can find details in Ref. \cite{pei2013spreading,lu2016vital,pei2018theories} and references therein.

The paper is organized as follows. In Section 2, the GC and k-core structure, on which the influencer identification problem is defined, are introduced. In this section, we discuss the links between regular percolation and independent cascade models, as well as the relationship between k-core percolation and threshold models. In Section 3, we discuss the progresses in optimal percolation using collective influence, optimal decycling, explosive percolation and large deviations of percolation in details. This section summarizes recent developments in finding the minimal set of nodes to collapse a network, i.e., the structural influencer identification problem. In Section 4, approaches for models with continuous transitions using greedy search and message passing are presented. In Section 5, methods developed for threshold models with discontinuous transitions are reported. Section 4 and Section 5 survey the methods to solve the dynamical influencer identification problem, with focus on dynamical models with continuous and discontinuous transitions respectively. Lastly, we conclude the review with an outlook of further directions in Section 6. 

\section{Giant component and k-core structure}

The topological feature of a network can be characterized by important structures such as the giant component and k-core. These concepts are fundamental in defining the problem of influencer identification in various dynamical processes. In this section, we introduce the regular percolation and k-core percolation processes, which are used to define the influencer identification problem, and elucidate their connections with commonly used spreading models.

\subsection{Percolation and independent cascade models}

\begin{figure}[]
  \centering
  \includegraphics[width=1\columnwidth]{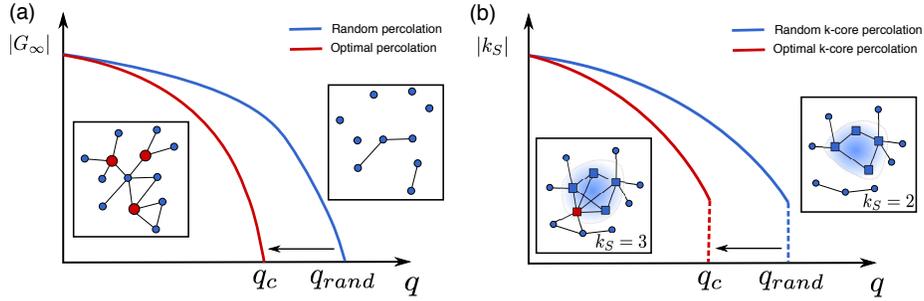}
  \caption{{Phase transition in percolation process.}
      (a). The continuous transition of the GC size $G_{\infty}$ after the removal of $q$ fraction of nodes. $q_{rand}$ and $q_c$ are the critical values for random and optimal percolation, respectively. Insets show illustrations of a connected $G_{\infty}$ and fragmented clusters. Red nodes are removed influencers. (b). The discontinuous transition of k-core size $|k_S|$ after the removal of $q$ fraction of nodes. The critical values for random and optimal k-core percolation are marked by $q_{rand}$ and $q_c$. Left inset shows the k-core of $k_S=3$. After the removal of the red node, 3-core collapses and only 2-core is left, as shown in the right inset.} \label{GCkcore}
      \end{figure}

The connectivity of a network is characterized by the number of nodes in the largest connected component, or the giant component (GC) $G_{\infty}$. In the random graph theory established by Paul Erd\H{o}s and Alfr\'ed R\'enyi in 1960s \cite{erdos1960evolution}, the percolation process describes the emergence of $G_{\infty}$ by gradually increasing the probability of connection between any pairs of nodes \cite{bollobas1998random}. In its inverse process, the giant component $G_{\infty}$ of an initially connected network collapses as an increasing fraction $q$ of nodes or links are removed. This removal process, termed site or bond percolation, leads to a continuous phase transition at a critical value of $q$, above which only fragmented clusters remain, as shown in Fig. \ref{GCkcore}(a).

For a network $G(V,E)$ with $N=|V|$ nodes and $M=|E|$ edges, we can use a vector $\bm{n}=(n_1,\cdots,n_N)$ to represent the configuration of whether a node $i$ is removed ($n_i=0$) or not ($n_i=1$).  After the removal of $q=1-\sum_{i=1}^Nn_i/N$ fraction of nodes, we define the size of remaining giant component $G_{\infty}(q)$ as the ratio of the number of nodes in $G_{\infty}$ to the network size $N$. In the classical percolation theory \cite{erdos1960evolution}, nodes are deleted randomly. At the critical value $q_{rand}$, $G_{\infty}$ is completely dismantled and becomes negligible compared with the network size $N$ in a continuous, or second order, phase transition. In thermodynamic limit $N\to\infty$, we have $\lim_{q\to q_{rand}^-}G_{\infty}(q)=0$ and $G_{\infty}(q)=0$ for $q\geq q_{rand}$. In real-world networks, the critical point $q_{rand}$ upon random attack depends on the heterogeneity of network structure. In particular, using generation functions, it was proved that for random networks with a given degree distribution $P(k)$, $q_{rand}$ is estimated by $\langle k\rangle/(\langle k^2\rangle-\langle k\rangle)$, where $\langle k\rangle=\sum kP(k)$ and $\langle k^2\rangle=\sum k^2P(k)$ are the first and second moments of $P(k)$ \cite{newman2001random}. This estimation predicts an extreme robustness to random attack, i.e., $q_{rand}\approx1$, for scale-free networks with a power law degree distribution $P(k)\propto k^{-\gamma}$, which are ubiquitous in real-world systems \cite{barabasi1999emergence,clauset2009power}. Later, more accurate estimations using the reciprocal of the largest eigenvalue of the adjacency matrix and non-backtracking matrix were developed \cite{radicchi2015predicting,karrer2014percolation}.

In random (or regular) percolation, nodes are removed without considering their difference in structural importance. As a matter of fact, if nodes are removed strategically, $G_{\infty}$ can be destructed well before the removal of $q_{rand}$ fraction of nodes. For example, $G_{\infty}$ in scale-free networks is extremely vulnerable to targeted attacks on hubs \cite{albert2000error,cohen2001breakdown,pastor2001epidemic}. The optimized process, deviating from the mean-field dynamics given by classical percolation theory, expedites the collapse of $G_{\infty}$ and reduces the critical value of $q$. In statistical physics and network science, a number of works have explored the large deviations of percolation., i.e. the deviations from the mean-field theory of percolation \cite{altarelli2013large,hartmann2011large,bianconi2018rare,bianconi2019large,coghi2018controlling}. At $q_{rand}$, there exist a number of possible configurations $\bm{n}$ such that $G_{\infty}(\bm{n})=0$. As $q$ decreases, fewer configurations satisfy $G_{\infty}(\bm{n})=0$, until at $q_c$ where only one configuration $\bm{n}^*$ exists. Below $q_c$, there is no solution to $G_{\infty}(q)=0$. Mathematically, $q_c=\min\{q\in[0,1] | G_{\infty}(q)=0\}$. The optimal percolation, or network dismantle problem is to find the unique configuration $\bm{n}^*$ corresponding to the minimal $q_c$, and influencers are the nodes with $n_i=0$ \cite{morone2015influence}. We note that the general framework of large deviations of percolation includes the optimal percolation problem as an extremal case \cite{altarelli2013large,bianconi2019large}.

By definition, percolation process concerns the pure structural integrity of networks. Nevertheless, a class of spreading dynamics can be mapped to percolation process and therefore studied using percolation theory. One of these dynamics is described by independent cascade models (ICMs) \cite{pastor2015epidemic,boccaletti2006complex,kempe2003maximizing}. In ICMs, an individual can be independently infected by any of his/her neighbors in the network. A spreading process starts from a set of infectious ``seeds'' in a susceptible population. In each time step, a susceptible individual can become infected by each of his/her infected neighbors with a certain transmission probability. Infected individuals keep infectious for a time of period, and then become susceptible again or immune to infection. The spreading process stops when there is no new infections. In applications, most widely used models include the susceptible-infected (SI) model, the susceptible-infected-susceptible (SIS) model and the susceptible-infected-removed (SIR) model \cite{hethcote2000mathematics,anderson1992infectious,diekmann2000mathematical}. These models are widely used in the simulation \cite{pastor2015epidemic,pastor2001epidemic,pastor2001epidemicpre,moreno2004dynamics,li2014rumor,yan2014dynamical}, detection \cite{altarelli2014bayesian,lokhov2014inferring,shah2010detecting,pei2015detecting,shah2011rumors,comin2011identifying} and forecast  \cite{pei2018forecasting,scarpino2019predictability,pei2017counteracting,kandula2018evaluation,pei2019predictability} of infectious disease spread and information diffusion. ICMs are closely related to percolation: the dynamical spreading process of an ICM can be transformed to a bond percolation with a given occupation probability \cite{newman2002spread}. As a result, the outcome of a dynamical ICM can be mapped to the static final state of an equivalent percolation process. This mapping bypasses the need to run dynamical models and enables us to analyze the spreading process using tools and properties of the well-studied percolation problem.

\subsection{K-core percolation and threshold models}

\begin{figure}[]
  \centering
  \includegraphics[width=0.5\columnwidth]{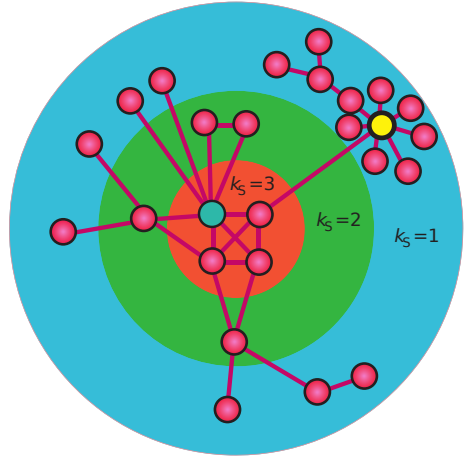}
  \caption{{An example of k-core decompostion.}
      The highlighted blue and yellow nodes have the same degree $k=8$, but with different $k_S$ values. Figure is adapted from \cite{kitsak2010identification} under permission from Springer Nature.} \label{kcore}
      \end{figure}

K-core decomposition classifies networks into layers with increasingly dense connections. In a network, the k-core is defined as the largest subgraph whose nodes have at least $k$ links \cite{seidman1983network,dorogovtsev2006k}. For example, the 1-core of a network is simply its GC; the 2-core is composed of all loops. Each node corresponds to a unique k-core index $k_S$ that indicates the highest k-core it locates. The k-core index $k_S$ is obtained through k-shell decomposition in which nodes are iteratively pruned according to their remaining degrees \cite{batagelj2011fast}. This process can be also viewed as a recursive calculation of the Hirsch-index $h$ \cite{lu2016h}, in which a node is assigned index $h$ if it has at least $h$ neighbors with degree no smaller than $h$ \cite{hirsch2005index}. Nodes with low $k_S$ values are located at the periphery of the network while the center consists of nodes with high $k_S$ values. An example of k-core decomposition is shown in Fig. \ref{kcore}. Recently, k-core percolation is generalized to multiplex networks \cite{azimi2014k}.

K-core structure provides higher-order information on network connectivity beyond giant component, which can be viewed as a 1-core. Originally proposed on lattices in statistical physics \cite{chalupa1979bootstrap}, k-core percolation (or bootstrap percolation) describes the formation process of k-core in networks \cite{goltsev2006k}. In a standard k-core percolation, nodes in a given network can be either active or inactive. Initially $p$ fraction of nodes are activated; in later steps, inactive nodes with at least $k$ active neighbors become activated recursively. In the final state, active nodes form the percolated k-core. The reversal process of k-core percolation depicts the destruction of k-core structure. Specifically, $q$ fraction of nodes are removed from the network, and nodes with less than $k$ neighbors are further recursively deleted. The final size of k-core $|k_S|$ is the fraction of nodes left.

K-core percolation has many important variants developed independently in other disciplines. For instance, in sociology, Granovetter proposed the threshold model of collective behavior in society in 1978 \cite{granovetter1978threshold}. In the well-studied version of linear threshold models (LTMs), nodes are activated only when the number of active neighbors exceeds a predefined threshold value. The heterogeneous k-core percolation, in which each node is assigned a local threshold, is a generalization of the classical k-core percolation and a special case of LTM \cite{cellai2011tricritical,baxter2011heterogeneous,cellai2013critical}. Within this framework, classical GC percolation can be viewed as a special case of LTM where the threshold of each node $i$ is $k_i-1$ ($k_i$ is the degree of node $i$) \cite{morone2015influence}.  Further, weights of interactions and nonlinear threshold rules have been introduced to describe more complex dynamics \cite{kempe2003maximizing,watts2002simple,dodds2004universal}. Recently, the k-core was also applied as a precursor of the jamming transition in granular materials \cite{morone2019jamming}.

More recently, a generalized k-core percolation was proposed as a generalization of the leaf removal process \cite{azimi2019generalization}. In this $k$-leaf removal algorithm, nodes of degree smaller than $k$ and their nearest neighbors together with all incident links are recursively pruned. The subgraph left after this pruning is called the Generalized $k$-core, or $Gk$-core. Similar as k-core percolation, the pruning procedure decomposes the network into layers of nested $Gk$-cores. However, as indicated by the authors, unlike k-core decomposition that classifies nodes according to their topological properties, the $Gk$-cores characterize a specific robustness of the network: it is actually the remained network after an epidemic that attacks weak individuals of degree less than $k$ and their neighbors.

The fundamental difference of k-core percolation from GC percolation is that the k-core size $|k_S|$ could undergo a discontinuous, or first order, phase transition under certain circumstances. For example, in Fig. \ref{GCkcore}(b), the left inset illustrates the 3-core of the network. Upon the removal of the red node, the 3-core is completely destroyed, with only 2-core left as shown in the right inset. In this example, the 3-core disappears abruptly from a non-zero size. Such discontinuous phase transition stems from the threshold rule of percolation, and lies at the heart of catastrophic cascading failures in many real-world systems \cite{buldyrev2010catastrophic,yang2017small}. A number of seminal works have explored the phase diagram and mechanism of transition in k-core percolation or threshold models \cite{baxter2010bootstrap,goltsev2006k,watts2002simple,dodds2004universal,dorogovtsev2006k-core,schwarz2006onset}. In particular, Watts modeled the global cascade on random networks using a linear threshold model and derived the critical condition for the discontinuous transition \cite{watts2002simple}. Goltsev {\it et al.} found the hybrid phase transition in k-core percolation with a discontinuous emergence of k-core as well as a continuous emergence of GC \cite{goltsev2006k}. In Ref. \cite{goltsev2006k}, authors demonstrated the crucial role of ``corona'', a subset of nodes in the k-core that have exactly $k$ neighbors: a random removal of even one node from the corona will trigger the collapse of a vast region of the k-core around the removed node. Baxter {\it et al.} further analytically derived the condition for the discontinuous transition of k-core in networks with arbitrary degree distributions \cite{baxter2010bootstrap}. The abrupt jump from a k-core with non-zero size to its collapse can be mathematically explained by a bifurcation of the dynamical system describing the k-core percolation. In such bifurcation, a small change of parameters (e.g., fraction of removed nodes) leads to the discontinuous shift of the stable point from a non-zero solution (k-core with non-zero size) to a zero solution (no k-core). Such bifurcation-induced transition is also responsible for the global cascade and vulnerability in interdependent networks and network of networks \cite{buldyrev2010catastrophic,parshani2010interdependent,gao2011robustness}.

Similar to optimal percolation, the configuration of node removal $\bm{n}$ can be optimized to induce early transition. For random k-core percolation, at the critical point $q_{rand}$, we have $\lim_{q\to q_{rand}^-}|k_S|>0$, $\lim_{q\to q_{rand}^+}|k_S|=0$ and $|k_S|(q)=0$ for $q>q_{rand}$. The optimal k-core percolation problem is to find the unique configuration $\bm{n}^*$ for which $q_c=\min\{q\in[0,1] | \lim_{q'\to q^+}|k_S|(q')=0\}$ (Fig. \ref{GCkcore}(b)). In some literature, this problem is also known as the minimal contagious set problem\cite{guggiola2015minimal,ackerman2010combinatorial,dreyer2009irreversible,reichman2012new,feige2017contagious,angel2017minimal,angel2016thresholds}. For threshold models, the influencer identification problem is to search for a given number of seeds that can lead to the maximal number of activated nodes.

\section{Optimal percolation}

Influence maximization is closely related to the optimal percolation problem. In addition, optimal percolation also provides a solution to the optimal immunization problem by dismantling the underlying network on which propagation occurs. Recently, within the message passing framework, Morone and Makse developed an efficient algorithm, the collective influence (CI), that gives a good approximation of optimal percolation \cite{morone2015influence}. Later, better algorithms based on optimal decycling \cite{braunstein2016network,mugisha2016identifying} and explosive percolation \cite{clusella2016immunization} were proposed. In this section, we discuss these structural approaches to the influence maximization problem.

\subsection{Collective influence}

Considering a network $G$ with $N$ nodes and $M$ edges, the vector $\mathbf{n}=(n_i,\cdots,n_N)$ encodes the configuration of whether node $i$ is removed ($n_i$=0) or not ($n_i$=1). Denoting the fraction of removed nodes by $q=1-\sum_{i=1}^Nn_i/N$, optimal percolation aims to find the minimal fraction $q_c$ of nodes such that the giant component $G_{\infty}$ is fully dismantled. Within the message passing framework, define message $\nu_{i\to j}$ as the probability that node $i$ belongs to $G_{\infty}$ without being connected to it through node $j$. Therefore, $\nu_{i\to j}=1$ if and only if $n_i=1$ and a least one of $i$'s neighbors other than $j$ is connected to $G_{\infty}$. For a locally tree-like structure, the messages evolves by the following equations:
\begin{equation}\label{percolationmp1}
\nu_{i\to j}=n_i\left [1-\prod_{m\in\partial i \setminus j}(1-\nu_{m\to i})\right ],
\end{equation}
where $\partial i\setminus j$ denotes the immediate neighbors of $i$ excluding $j$. Taking node $j$ back into consideration, the probability that $i$ is connected to the giant component is then calculated as
\begin{equation}\label{percolationmp2}
\nu_{i}=n_i\left [1-\prod_{m\in\partial i}(1-\nu_{m\to i})\right ].
\end{equation}

By linearizing Eq. (\ref{percolationmp1}) around the fixed point $\{\nu_{i\to j}=0\}$, the stability of this solution is determined by the largest eigenvalue $\lambda(\mathbf{n};q)$ of a linear operator $\mathcal{M}$. Specifically, $\mathcal{M}$ is the Jacobian of the system defined on $2M\times 2M$ directed edges as $\mathcal{M}_{m\to n,i\to j}\equiv \frac{\partial\nu_{i\to j}}{\partial\nu_{m\to n}}|_{\{\nu_{i\to j}=0\}}$. A few calculations show that the matrix $\mathcal{M}$ can be represented in terms of the non-backtracking (NB) matrix $\mathcal{B}$ \cite{hashimoto1989zeta} via
\begin{equation}\label{MNB}
\mathcal{M}_{m\to n,i\to j}=n_i\mathcal{B}_{m\to n,i\to j},
\end{equation}
where the NB matrix is
\begin{equation}
\mathcal{B}_{m\to n,i\to j}=\left\{
\begin{aligned}
1 & \text{ if } n=i \text{ and } j\neq m, \\
0 & \text{ otherwise.} \\
\end{aligned}
\right.
\end{equation}
The matrix entry $\mathcal{B}_{m\to n,i\to j}$ is non-zero only when ($m\to n$, $i\to j$) form a pair of consecutive non-backtracking directed edges, i.e., ($m\to n$, $n\to j$) with $m\neq j$. For non-backtracking edges, $\mathcal{B}_{m\to n,n\to j}=1$.

Following the Frobenius theorem, the largest eigenvalue $\lambda(\mathbf{n};q)$ is real and positive. The solution $\{\nu_{i\to j}=0\}$ is stable if $\lambda(\mathbf{n};q)\leq1$. In this way, the optimal percolation problem can be solved by finding the optimal configuration $\mathbf{n}^*$ such that $\lambda(\mathbf{n}^*;q_c)=1$. For $q<q_c$, all configurations lead to $\lambda(\mathbf{n};q)>1$. On the contrary, for $q>q_c$, there exist two different circumstances. For some non-optimal configurations, the macroscopic component still exists. On the other hand, there are also configurations such that $\lambda(\mathbf{n};q)\leq1$, which correspond to a fully fragmented network. As $q\to q_c^+$, the number of configurations satisfying $\lambda(\mathbf{n};q)\leq1$ gradually decreases and eventually vanishes at $q_c$, where the optimal configuration $\mathbf{n}^*$ is obtained. To develop a scalable algorithm, the eigenvalue can be approximated using the Power Method \cite{saad2011numerical}. For a given number of iterations $\ell$, the collective influence (CI) of node $i$ can be defined as:
\begin{equation}\label{CIequation}
\text{CI}_\ell(i)=(k_i-1)\sum_{j\in\partial\text{Ball}(i,\ell)}(k_j-1),
\end{equation}
where $\partial\text{Ball}(i,\ell)$ is the frontier of the ball of radius $\ell$ in terms of shortest path centered around node $i$. By iteratively removing the node with largest CI, the largest eigenvalue of $\mathcal{M}$ can be minimized with high efficiency. After each removal, the CI score of every remaining node in the network is recalculated. This process continues until the network is fully fragmented, i.e. $G_{\infty} \ll 1$. The optimal configuration $\mathbf{n^*}$ and $q_c$ are estimated from this removal process.

For $q<q_c$, the network can not be fully dismantled. In order to obtain the smallest giant component, a greedy reinsertion procedure is performed starting from the optimal configuration $\mathbf{n}^*$. In the reinsertion procedure, an index $c(i)$ is define for each removed node. Specifically, $c(i)$ is the number of clusters that would be joined together if node $i$ is put back in the network. Nodes with the smallest $c(i)$ score are iteratively reinserted until the fraction of removed nodes decreases to $q$.

\begin{figure}[]
  \centering
  \includegraphics[width=1\columnwidth]{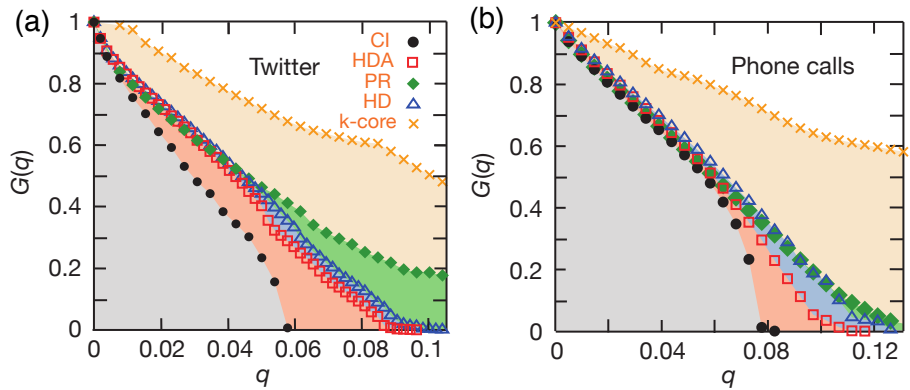}
  \caption{{Performance of collective influence in large-scale real social networks.}
       Giant components $G(q)$ for a Twitter network (a) and a social network of mobile phone users in Mexico (b) computed using CI (collective influence), HDA (high degree adaptive), PR (PageRank), HD (high degree) and k-core strategies are compared. Figure is reused from \cite{morone2015influence} permitted by Springer Nature.} \label{CI}
      \end{figure}

The computational complexity of the CI algorithm is $O(N^2)$. In practice, it can be accelerated by limiting the calculation and update of CI inside the $(\ell+1)$-ball around the removed node. In addition, the complexity can be further reduced to $O(N\log N)$ by sorting the CI scores in a heap structure \cite{morone2016collective}, which makes it scalable to large networks. Simulation results on both synthetic and real-world social networks show that the CI algorithm outperforms the equal graph partitioning (EGP) immunization strategy \cite{chen2008finding} and frequently used heuristic metrics such as degree centrality, PageRank and k-core index. For a Twitter network with $469,013$ users and a Mexico mobile communication network of $1.4\times 10^7$ users, the CI algorithm achieves fully fragmentation with a smaller set of influencers \cite{morone2015influence} (see Fig. \ref{CI}). For such massively large-scale networks, a variant of the CI algorithm can be applied without losing performance by removing a finite fraction of nodes instead of one node at each step.

Given a finite radius $\ell$, the CI algorithm is local in nature. To incorporate the influence of a node at the global level, Morone {\it et al.} improved the CI algorithm using a message-passing approach and proposed the CI propagation algorithm ($\text{CI}_\text{P}$) \cite{morone2016collective}. As a variant of the CI algorithm in the limit of $\ell\to\infty$, the $\text{CI}_\text{P}$ algorithm is able to reach the analytical optimal percolation threshold of random cubic graphs \cite{bau2002decycling}. Another belief-propagation variant of CI algorithm, $\text{CI}_\text{BP}$, was also proposed. Combining the dynamics of the SIR model with message-passing updating rules, $\text{CI}_\text{BP}$ achieves similar performance with $\text{CI}_\text{P}$. However, the improvements of $\text{CI}_\text{P}$ and $\text{CI}_\text{BP}$ over CI are made at the expense of increasing computational complexity from $O(N\log N)$ to $O(N^2\log N)$. Kobayashi and Masuda recently developed an immunization algorithm for networks with community structure combining the CI algorithm and coarse graining procedure in which communities were regarded as supernodes \cite{kobayashi2016fragmenting}. From a mesoscopic scale, nodes connecting different communities can be identified at a cost of $O((N^2/N_C)\log N)$ ($N_C$ is the number of communities). The optimal percolation problem was also studied on multiplex networks. Osat {\it et al.} showed that characteristics in multiplex networks such as edge overlap and interlayer degree-degree correlation could profoundly change the properties of influencers \cite{osat2017optimal}. Neglecting the multiplex structure of a network would lead to significant inaccuracies about its robustness. In applications, the collective influence theory has been used to locate superspreaders of information in real-world social media \cite{teng2016collective}, find sources of fake news in Twitter during the 2016 US presidential election \cite{bovet2018validation,bovet2019influence}, single out critical regions in brain networks \cite{del2018finding,morone2017model}, infer personal economic status \cite{luo2017inferring}, improve cooperation in evolutionary games \cite{szolnoki2016collective} and control biological networks \cite{zhang2018dynamic,wang2018optimal,wang2019on,pei2012how}.

As demonstrated in Ref. \cite{morone2015influence}, the optimal percolation problem can be mapped exactly onto the influence maximization problem for the linear threshold model with threshold $k_i-1$ ($k_i$ is the degree of node $i$). As a result, the CI algorithm, designed for optimal percolation, also provides a solution to the influence maximization problem for this specific transmission model. For linear threshold models with other threshold values, the CI algorithm was generalized to solve the influence maximization problem with first-order transitions, which will be addressed later. In addition, a detailed discussion on the relation of the CI algorithm with the SIR model can be found in Ref. \cite{morone2015influence}.

\subsection{Optimal decycling-based algorithms}

Recent works have shown that the optimal percolation problem is closely related to the optimal decycling problem, or minimum feedback vertex set (FVS) problem \cite{braunstein2016network,mugisha2016identifying}. A feedback vertex set is the set of nodes whose removal would break all the loops in the network \cite{karp1972reducibility}. The optimal decycling problem is, in fact, analogous to find the FVS with smallest number of nodes. The rationale behind the connection between optimal percolation and optimal decycling is that, for sparse random networks, short loops rarely exist in small connected components \cite{marinari2004circuits,marinari2005algorithm,bianconi2005loops}. If the long loops in the giant component are cut, the network will break into small tree fragments. As indicated by Braunstein {\it et al.} \cite{braunstein2016network}, the optimal decycling threshold $q_c^{dec}$ acts as an upper bound of the optimal percolation threshold $q_c$. For random networks with light-tailed degree distribution (finite second moment), the minimal size of decycling set is equal to the minimal size of dismantling set in the limit $N\to\infty$. The optimal decycling problem is itself an NP-hard problem, but can be solved via belief propagation algorithms approximately. Two approaches based on decycling algorithm were developed recently \cite{braunstein2016network,mugisha2016identifying}. Both of them apply a three-stage procedure: first decycle the network with minimal number of nodes, then break the tree into small components, and finally reinsert some nodes to the network without increasing the size of the largest component. Compared with the CI algorithm, these two algorithms take into account the global topology of the network and achieve a better performance.

The belief-propagation-guided decimation (BPD) algorithm proposed by Mugisha and Zhou is based on the spin glass model of the FVS problem \cite{zhou2013spin}. In order to transform the global acyclic constraint into local ones, a variable $A_i$, which takes the value $0$, $i$ or $j\in \partial i$, is assigned to each node \cite{zhou2013spin}. If node $i$ is removed from $G$, $A_i=0$. Otherwise, $A_i=i$ if it is a root of a tree or $A_i=j$ if node $i$ has a parental node $j$. Given a microscopic configuration $\mathbf{A}=\{A_1,A_2,\cdots,A_N\}$, the fraction of removed nodes is represented by:
\begin{equation}
q^{dec}(\mathbf{A})=\frac{1}{N}\sum_{i=1}^N\delta_{A_i}^0,
\end{equation}
where $\delta_n^l$ is the Kronecker delta function ($\delta_n^l=1$ if $n=l$ and 0 otherwise). For each edge $(i,j)$ in the network, an edge factor $C_{ij}(A_i,A_j)$ is defined as \cite{zhou2013spin}:
\begin{eqnarray}
C_{ij}(A_i,A_j) &=&\delta_{A_i}^0\delta_{A_j}^0+\delta_{A_i}^0(1-\delta_{A_j}^0-\delta_{A_j}^i)+\delta_{A_j}^0(1-\delta_{A_i}^0-\delta_{A_i}^j)\nonumber\\
&&+\delta_{A_i}^j(1-\delta_{A_j}^0-\delta_{A_j}^i)+\delta_{A_j}^i(1-\delta_{A_i}^0-\delta_{A_i}^j).
\end{eqnarray}
The edge factor $C_{ij}(A_i,A_j)$ is either 1 or 0. The edge $(i,j)$ is regarded as satisfied if $C_{ij}(A_i,A_j)=1$, and unsatisfied otherwise. For a configuration $\mathbf{A}$, if all edges in a network $G$ are satisfied, we define $\mathbf{A}$ as a solution of $G$. The definition of satisfied edges relaxes the original problem of acyclic components to allow at most one cycle in the remained components. Indeed, it has been proven that all remaining nodes in a graph for a solution $\mathbf{A}$ form a subgraph consisting of several components that each contains at most one cycle. Considering all solutions of the network, a partition function of the system is defined as:
\begin{equation}
Z(\mu)=\sum_{\mathbf{A}}e^{\mu N(1-q^{dec}(\mathbf{A}))}\prod_{(i,j)\in G}C_{ij}(A_i,A_j),
\end{equation}
where $\mu$ is the inverse of temperature. At the limit of zero temperature, the partition function is contributed exclusively by the optimal configuration $\mathbf{A}^*$ with the minimal fraction $q_c^{dec}$.

Under locally tree-like assumption, the marginal probability $q_i^0(t)$ for node $i$ to be removed from the remaining network $G(t)$ can be calculated through iterations of a set of belief propagation (BP) equations \cite{mugisha2016identifying}. At each time step $t$, the BP equations are iterated for a given number of rounds and the removal probability $q_i^0$ is calculated for each node. The node with the highest probability $q_i^0$ is removed from the network even if the BP equations do not converge to a fixed point. The process stops when the network becomes acyclic. If the largest component $G_{\infty}$ remains extensive, it can be further fragmented by iteratively deleting nodes that lead to the smallest giant component. The BPD algorithm can be well applied to networks with rare short loops. However, for a large number of networks with abundant communities, the nodes in FVS set are usually more than necessary to dismantle the network stricture. Therefore a reinsertion process can be proceeded without significantly increasing the size of $G_{\infty}$. This process can be done through a greedy algorithm, in which the nodes that cause the least increase in $G_{\infty}$ are reinserted one after another until the size of $G_{\infty}$ reaches a predefined threshold.

\begin{figure}[]
  \centering
  \includegraphics[width=1\columnwidth]{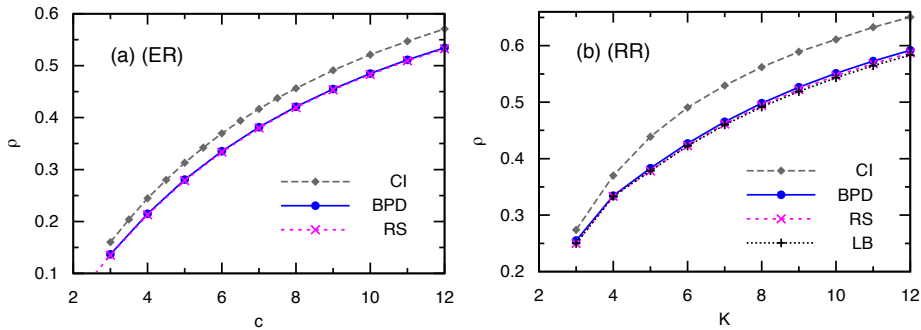}
  \caption{{Performance of the BPD algorithm in ER and RR networks.}
      Fraction of removed nodes $\rho$ to break ER random networks of mean degree $c$ (a) and regular random networks of degree $K$ (b). Diamonds show the results obtained from collective influence (CI). Crosses represent results of the replica-symmetric (RS) mean-field theory. Plus symbols in (b) show the mathematical lower bound (LB) on the minimum size of target nodes. Figure reuse from \cite{mugisha2016identifying} is permitted by American Physical Society.} \label{BPD}
      \end{figure}

The BPD algorithm is scalable to large networks with a computational complexity of $O(N\log N)$. Simulations on random network ensembles and real-world networks indicate that the BPD algorithm is superior to the CI algorithm in optimal percolation problem (see Fig. \ref{BPD}). However, as shown in Ref. \cite{morone2016collective}, the BPD algorithm is relatively slower than the CI algorithm. For large random Erd\H{o}s-R\'enyi (ER) networks and scale-free networks, the BPD algorithm manages to fragment the network by removing a smaller set of nodes compared with CI algorithm. In particular, the percolation threshold is close to the minimal value predicted by the replica-symmetric (RS) mean field theory \cite{bau2002decycling,zhou2013spin}. In the CI algorithm, the size of $G_{\infty}$ decreases almost linearly with the increase of $q$. In contrast, $G_{\infty}$ features an abrupt collapse under the BPD algorithm. This results from the intrinsic nature of the FVS problem and the efficiency of tree dismantling. With the existence of such collapse, the BPD process can work as an efficient attack strategy, leaving no warning to the system before its total failure. In a recent work on dismantling efficiency and network fractality \cite{im2018dismantling}, it was found that the BPD algorithm outperforms the CI algorithm no matter whether the network is fractal or not, while the CI algorithm works better on non-fractal networks, which have high ratios of long-range shortcuts to short-range connections.

Braunstein {\it et al.} considered the optimal decycling problem from a different point of view \cite{braunstein2016network}. In this work, the optimal percolation problem was named as network dismantle. Noticing that a network is acyclic if and only if its 2-core is empty, authors mapped the decyling process to a 2-core percolation. Assume a set of nodes $S\subset V$ are initially removed from the network. The 2-core percolation can be described by the evolution of time-dependent binary variables $x_i^t(S)$ for $1\leq i\leq N$. Starting from the initial setting $x_i^0(S)=1$ for removed nodes $i\in S$ and $x_i^0(S)=0$ for $i\notin S$ at $t=0$, the evolution follows \cite{braunstein2016network}
\begin{equation}x_i^{(t+1)}=
\begin{cases}
1 & \text{if }x_i^t(S)=1\\
\mathbb{I}\left[\sum_{j\in \partial i}\left(1-x_j^t(S)\right)\leq 1\right] & \text{if }x_i^t(S)=0
\end{cases}
\label{2core}
\end{equation}
where the indicator function $\mathbb{I}$ is 1 if the argument is true and 0 otherwise. As $x_i^t$ can only change from 0 to 1,  the equations admit a fixed solution $x_i^*(S)$ as $t\rightarrow \infty$. In particular, $x_i^*(S)=0$ iff $i$ belongs to the 2-core of $G\setminus S$. If $x_i^*(S)=1$ for all nodes, $G\setminus S$ contains no loops and $S$ is called a decycling set. To find the minimal decycling set, it is convenient to introduce the probability distribution over decycling sets $S$ using the Boltzmann distribution in statistical physics
\begin{equation}
\hat{\eta}(S)=\frac{1}{Z(\mu)}e^{\mu |S|}\prod_{i\in V}\mathbb{I}\left[x_i^*(S)=1\right],
\end{equation}
where $|S|$ is the number of nodes in $S$, $\mu$ is the inverse temperature, and $Z(\mu)$ is the partition function that normalizes the distribution. The minimal size of decycling sets can be calculated in the zero-temperature limit: $q_c^{dec}=\lim_{\mu\rightarrow -\infty}\frac{1}{N\mu}\ln Z(\mu)$.

Since $x_i^*$ depends on $S$ in a global way, it is difficult to compute $Z(\mu)$ directly. To solve this problem, authors transformed the global constraint $\prod_{i\in V}\mathbb{I}\left[x_i^*(S)=1\right]$ to its local equivalent. The node removal process in 2-core percolation can be described by an integer $t_i(S)=\min\{t:x_i^t(S)=1\}$ defined for each node $i$, which encodes the time when node $i$ is removed from the network. For $i\in S$, it is straightforward that $t_i(S)=0$. For $i\notin S$, $t_i(S)$ depends locally on its neighbors according to
\begin{equation}
t_i(S)=\phi_i(\{t_j\}_{j\in \partial i})=1+\text{max}_2(\{t_j(S)\}_{j\in \partial i}),
\end{equation}
where the function $\max_2$ returns the second largest value in its argument. Under this parameterization, the partition function can be rewritten as
\begin{equation}
Z(\mu)=\sum_{\{t_i\}}e^{\mu\sum_i\psi_i(t_i)}\prod_{i\in V}\mathbb{I}[t_i<\infty]\mathbb{I}\left[t_i=\phi_i(\{t_j\}_{j\in\partial i})\right],
\label{partition}
\end{equation}
where $\psi_i(t_i)=\mathbb{I}[t_i=0]$.

The exact computation of Eq. (\ref{partition}) is NP-hard. In calculation, a simplification of the partition function in Eq. (\ref{partition}) can be performed by restricting $t_i$ to be no larger than $T$. All values $t_i$ larger than $T$ are regarded as infinity. Under this simplification, trees with diameters larger than $T+1$ are considered to be part of a long cycle. Given a large enough $T$, the effect of this simplification is negligible. For locally tree-like graphs, the partition function can be computed by the cavity method \cite{mezard2009information,mezard2001bethe}, in which ``messages'' are exchanged between neighboring nodes. For each link $i\to j$, a message $\eta_{ij}(t_i,t_j)$ as a function of activation times $t_i$ and $t_j$ is introduced. The messages satisfy the self-consistent BP equations \cite{braunstein2016network}:
\begin{equation}
\eta_{ij}(t_i,t_j)\propto \sum_{\{t_k\}_{k\in \partial i\setminus j}}e^{\mu\psi_i(t_i)}\mathbb{I}\left[t_i=\phi_i(\{t_k\}_{k\in\partial i})\right]\prod_{k\in \partial i\setminus j} \eta_{ki}(t_k,t_i).
\end{equation}

As the temperature approaching zero ($\mu\to -\infty$), probabilities of the messages $\eta_{ij}(t_i,t_j)$ in the BP equations concentrate on the solution to Eq. (\ref{2core}) that minimizes the cost function $\sum_i\psi_i(t_i)$. To develop an algorithm that finds the optimal decycling set, a slightly different cost function is used: $\psi_i(t_i)=\mathbb{I}[t_i=0]+\varepsilon_i(t_i)$, where $\varepsilon_i(t_i)$ is a randomly chosen small cost. Further, the 2-core percolation process is relaxed to allow $t_i\geq 1+\max_2(\{t_j\}_{j\in\partial i})$ in Eq. (\ref{2core}). Define $h_i(t_i)$ as the minimal cost to dismantle the 2-core under the condition that node $i$ is removed at $t_i$. The optimal decycling set is determined by $S^*=\{i \in V| t_i^*=0\}$, where $t_i^*=\arg\min h_i(t_i)$. In calculation, $h_i(t_i)$ can be computed using Min-Sum algorithm, which is derived at the zero temperature limit of BP equations. Concretely, messages $h_i(t_i)$ are solved by iterating a set of equations \cite{braunstein2016network}. In most cases, convergence can be reached within a small number of iterations, with a computational complexity $O(MT)$ in each iteration. In case the Min-Sum equations do not converge, a reinforcement procedure is applied to damp the system \cite{bayati2008statistical}. 

In the acyclic network $G\backslash S^*$, there may still exist some extensive tree components. These large trees can be fragmented efficiently via a greedy tree breaking procedure with computational complexity of $O(N(\log N+T))$. In addition, for networks containing many short loops, a reverse greedy (RG) reinsertion procedure is applied to recover the nodes that do not increase the size of the giant component, as performed in the CI and BPD algorithms. The computational cost of this RG procedure is $k_{max}C'\log(k_{max}C')$, where $k_{max}$ is the maximal degree and $C'$ is the upper bound of $G_{\infty}$ size.

\begin{figure}[]
  \centering
  \includegraphics[width=0.8\columnwidth]{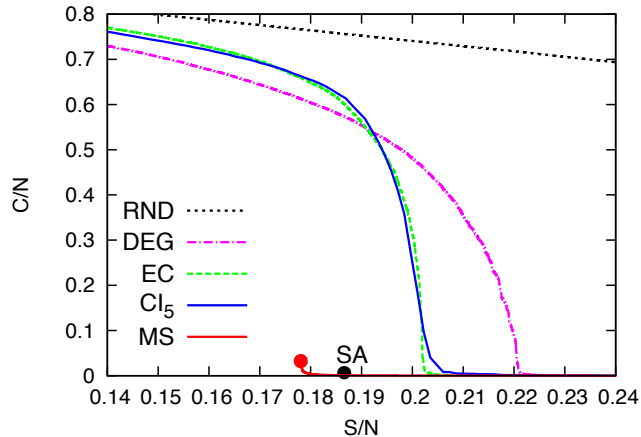}
  \caption{{Performance of the Min-Sum algorithm in an ER network.}
       Relative size of the largest component $C/N$ as a function of the fraction of nodes $S/N$ removed from an ER network (size $N=78,125$ and average degree $d=3.5$). Comparisons are performed for the Min-Sum algorithm (MS), random (RND), adaptive largest degree (DEG), adaptive eigenvector centrality (EC), adaptive CI, and simulated annealing (SA). Figure reuse from \cite{braunstein2016network} is permitted by National Academy of Sciences.} \label{ND}
      \end{figure}

Simulations on both synthetic and real-world social networks demonstrate the effectiveness of this decycling based algorithm. For an ER random graph of size $N=78,125$ and average degree $d=3.5$, the $G_{\infty}$ size deceases to $0.032$ when $17.81\%$ of nodes are removed. Compared with metrics of degree centrality, eigenvector centrality and the CI algorithm with $\ell=5$, it was found that the three-stage algorithm is superior in dismantling the giant component (see Fig. \ref{ND}). The Monte Carlo-based simulated annealing (SA) algorithm gives a competitive result. However, its computational complexity is much higher. For the same Twitter network analyzed in Ref. \cite{morone2015influence}, the Min-Sum algorithm with RG performs equally well with SA, removing only $3.4\%$ of nodes to break the giant component (smaller than 1,000 nodes). In comparison, CI needs to remove $4.6\%$ of nodes to achieve the same fragmentation performance.

Inspired by the decycling-based algorithm, a simple and faster heuristic algorithm with complexity $O(N)$, CoreHD, was developed \cite{zdeborova2016fast}. Starting from the 2-core of a network $G$, CoreHD recursively removes nodes with the highest degree in the 2-core until $G$ is fully dismantled. Despite its simpleness, CoreHD is reported to perform better than the CI algorithm. Specially, for large random networks, the performance of CoreHD is close to the theoretical solution predicted by replica-symmetry and 1RSB approximation \cite{guggiola2015minimal}. In addition, this simple algorithm is amenable to rigorous analysis, performing well even on loopy networks which are not accessible for typical message-passing algorithms. In a recent work by Schmidt {\it et al.} \cite{schmidt2019minimal}, the CoreHD algorithm was analyzed rigorously by translating the node removal in the CoreHD algorithm to a random process on the degree distribution of the network. The mapped dynamics, described by a set of coupled nonlinear ordinary differential equations, characterize the behavior of the CoreHD algorithm on random graphs. In the analysis, new upper bounds on the size of the minimal contagious sets in random graphs were proposed, which improves the best known results \cite{guggiola2015minimal,bau2002decycling}. The CoreHD analysis also inspired an improved heuristic algorithm, WEAK-NEIGHBOR, that works for both optimal percolation and k-core percolation \cite{schmidt2019minimal}. Details of this algorithm will be introduced in the next section.

\subsection{Explosive percolation-based immunization}

Another approach of optimal percolation was developed by Clusella {\it et al.} \cite{clusella2016immunization} based on explosive percolation (EP). In contrast with ordinary bond percolation which usually exhibits second or higher order phase transitions, EP features an unusual threshold behavior -- an explosive emergence of the giant component at the critical point 
\cite{achlioptas2009explosive,da2010explosive,riordan2011explosive,grassberger2011explosive,friedman2009construction}. To obtain an explosive transition, Achlioptas {\it et al.} proposed a modified edge addition procedure, wherein, at each step, two candidate edges are chosen randomly, but only one of them is actually occupied \cite{achlioptas2009explosive}. Given the weight of a node measured by the size of the connected component it belongs to, the edge with the minimal sum or product of nodes' weights is selected. These two procedures are referred to as the min-cluster and min-product rule. Compared with the random occupation of edges in ordinary percolation, the min-cluster or min-product rule favors the connection between small components, hereby suppresses the generation of an extensive component.

The explosive immunization (EI) algorithm adopts an inverse strategy that starts from a configuration where all nodes are virtually removed ($q=1$). Then less ``dangerous'' nodes are progressively unvaccinated. The procedure is performed in two schemes for $q>q_c$ and $q<q_c$, each of which uses a score to rank nodes in terms of their suitability to be unvaccinated. Similar to the construction of EP, in each time step, $m$ candidates (typically $m\approx 10^3$) are randomly selected. For $q>q_c$, the node with the lowest blocking ability (the weakest blocker) is put back into the network. The blocking ability is quantified by a score $\sigma_i^{(1)}$, which is a synthesis of the size of clusters it would join and its local effective connectivity. Specifically, the score $\sigma_i^{(1)}$ is defined as \cite{clusella2016immunization}: $\sigma_i^{(1)}=k_i^{(\text{eff})}+\sum_{\mathcal{C}\subset\mathcal{N}_i}(\sqrt{|\mathcal{C}|}-1)$, where $\mathcal{N}_i$ is the set of all components connected to node $i$ and $|\mathcal{C}|$ is the size of a component $\mathcal{C}$. $k_i^{(\text{eff})}$ measures the ``effective'' connectivity of node $i$ based on the local structure of its neighborhood and can be determined by a set of closed equations \cite{clusella2016immunization}: $k_i^{(\text{eff})}=k_i-L_i-M_i(\{ k_j^{(\text{eff})}\}_{j \in \partial i})$, where $k_i$ is the degree of node $i$, $L_i$ is the number of leaves in the neighborhood of node $i$ and $M_i$ returns the number of strong hubs. The strong hubs are defined recursively as nodes with $k_i^{(\text{eff})}$ larger than a threshold value (set as 6 in applications). The terms $L_i$ and $M_i$ are subtracted from $k_i$ since leaves have no contribution to connectivity and hubs are more likely to be removed in explosive immunization. 

In the first part of the EI algorithm, the node with the lowest $\sigma_i^{(1)}$ score among $m$ candidates is unvaccinated in each iteration. This process eventually reaches a critical fraction of immunized nodes $q_c$ where the $G_{\infty}$ size exceeds a small threshold value. In the region of $q<q_c$, however, the same procedure will lead to an abrupt jump of the $G_{\infty}$ size when two large components are joined together. As a consequence, in the second part at $q<q_c$, another score $\sigma_i^{(2)}$ is used to suppress such explosive growth of the giant component. The definition of $\sigma_i^{(2)}$ reads \cite{clusella2016immunization}
\begin{equation}
\sigma_i^{(2)}=
\begin{cases}
\infty               & \text{if } G_{\infty} \subsetneq\mathcal{N}_i, \\
|\mathcal{N}_i|      & \text{else, if arg min}_i |\mathcal{N}_i| \text{ is unique,}\\
|\mathcal{N}_i|+\epsilon|\mathcal{C}_2|      & \text{else},
\end{cases}
\end{equation}
where $|\mathcal{N}_i|$ is the number of components connected to $i$, $\mathcal{C}_2$ is the second largest component in $\mathcal{N}_i$, and $\epsilon$ is a small positive number. According to the score $\sigma_i^{(2)}$, the selection is made only among the neighborhood of $G_{\infty}$. The candidate with the smallest number of neighboring components is favored; if it is not unique, the one with the smallest $|\mathcal{C}_2|$ is selected. This process is proceeded recursively until the fraction of vaccinated nodes $q$ reaches the expected value.

\begin{figure}[]
  \centering
  \includegraphics[width=0.6\columnwidth]{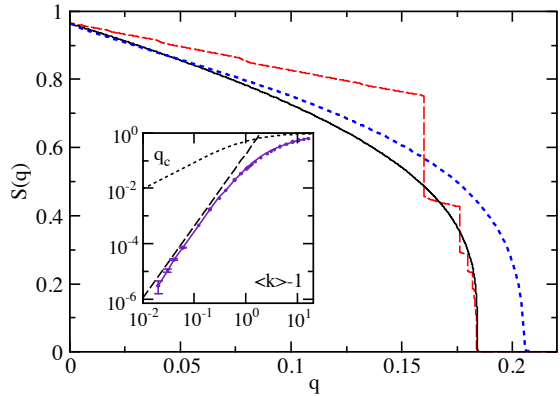}
  \caption{{Performance of the EI algorithm in an ER network.}
      Relative size of the largest clusters $S(q)$ after $q$ fraction of nodes are removed from ER networks with size $N=10^6$ and average degree $\langle k\rangle=3.5$. The red dashed curve is obtained by EI with score $\sigma^{(1)}$. The black continuous curve is obtained with $\sigma^{(2)}$ for $q<q^*$ ($S(q^*)=1/500$). The blue dotted line is the result of CI. The inset shows the relationship between $q_c$ and $\langle k\rangle-1$. The dotted line is the result for random strategy. Figure reuse from \cite{clusella2016immunization} is permitted by American Physical Society.}\label{Explosive}
      \end{figure}

Using the Newman-Ziff percolation algorithm in identifying susceptible components \cite{newman2001fast}, the explosive immunization algorithm is computationally efficient, which scales as $O(N\log N)$. In addition, it can be accelerated further by considering a small number of candidates. Simulations on both synthetic and real-world networks indicate that the explosive immunization algorithm outperforms the CI algorithm (see Fig. \ref{Explosive}). As a matter of fact, it achieves the smallest percolation threshold $q_c$ except for the belief propagation algorithms in Ref. \cite{braunstein2016network,mugisha2016identifying}.

\subsection{Graph partition-based algorithm}

In an earlier work, the optimal immunization problem was solved by an equal graph partitioning (EGP) immunization strategy based on the heuristic optimal partitioning of graphs \cite{chen2008finding}. In EGP, the network is fragmented into small connected clusters of approximately equal size. In a targeted attack on high-degree nodes, clusters after fragmentation have a broad distribution of sizes, including many small clusters. The targeted strategy may select high-degree nodes in these small clusters, which are unnecessary in breaking down the network. The EGP method avoids fragmenting small clusters, as the clusters all have similar sizes. In the EGP method, a network is first separated into two components with arbitrary size ratio by a minimal number of separators, solved using the nested dissection (ND) algorithm \cite{lipton1979generalized}. Then the network can be partitioned into any desirable number of same size clusters by applying ND algorithm recursively. This greedy graph-partitioning strategy provides $5\%$ to $50\%$ improvement over the targeted strategy on model networks and real-world networks.

The original network dismantle problem was recently extended to a generalized network dismantle problem in which the cost of removing a node is considered \cite{ren2019generalized}. In real-world systems, attacking important nodes typically requires a high cost as they are usually well protected. The generalized network dismantle problem seeks to find a set of nodes whose removal would fragment a network at the minimal cost.

Authors solved this problem by recursively applying node-weighted partition, i.e., partition a network into two parts of same size by removing a minimal number of edges. Specifically, define $v_i=+1$ if node $i$ belongs to a subgraph $M$ and $v_i=-1$ if node $i$ belongs to its complement $\bar{M}$. Assuming that the cost of cutting a link $(i, j)$ equals the cost of removing nodes $i$ and $j$, a node-weighted spectral cut objective function was proposed \cite{ren2019generalized}:
\begin{equation}
\frac{1}{2}\sum_{i,j}-\frac{1}{2}(v_iv_j-1)A_{i,j}(w_i+w_j-1),
\end{equation}
where $A$ is the adjacency matrix, and $w_i$ is the cost of removing node $i$. The optimization problem was then written in matrix notation as minimizing $v^TL_wv/4$ subject to $\sum_i v_i=0$,  where $L_w=D_B-B$ is the node-weighted Laplacian of the matrix $B=AW+WA-A$ ($W$ and $D_B$ are diagonal matrices with elements $W_{ii}=w_i$ and $(D_B)_{ii}=\sum_j B_{ij}$.)

The problem with integer constraint $v_i\in\{+1,-1\}$ is difficult to solve. As a result, the problem is relaxed to allow a real number $v_i\in\mathbf{R}$. For the relaxed problem, the solution of $v$ is analytically given by the second-smallest eigenvector of $L_w$, denoted by $v^{(2)}$. To approximate this solution, the matrix $L_w$ is transformed so that $v^{(2)}$ becomes the second-largest eigenvector. The eigenvector problem is solved by power iteration, with the initial vector set perpendicular to the largest eigenvector of the transformed matrix. Once $v^{(2)}$ is obtained, the separating edges are those connecting nodes with $v_i\geq 0$ to nodes with $v_i<0$. The set of nodes to be removed are optimized to cover all separating edges with minimal cost, which is transformed to the weighted vertex cover problem \cite{bar1981linear}. Finally, a reinsertion procedure is applied to find the nodes that are not necessary to fragment networks.

The generalized network dismantle (GND) algorithm has complexity $O(N\log^{2+\epsilon}(N))$, which can be applied to large-scale networks. For nonunit costs, the GND algorithm outperforms current state-of-the-art; for unit cost, it performs better than or comparable to state-of-the-art \cite{ren2019generalized}.

\subsection{Large deviations of percolation}

The optimal percolation problem can be studied within the framework of large deviations of percolation. Generally, in the BP equations that describe the percolation process, the {\it inverse} temperature $\beta$ in the Boltzmann distribution of configurations $\mathbf{n}$, $e^{-\beta\mathcal{E(\mathbf{n})}}$ ($\mathcal{E(\mathbf{n})}$ is energy defined by the size of giant component for $\mathbf{n}$), controls the deviation of dynamics from random percolation. For instance, an infinity temperature ($\beta=0$) corresponds to the random scenario, where each configuration is equally possible. As the temperature decreases, the dynamics start to deviate from the random scenario to more extreme cases: the distribution of configurations will concentrate on rare configurations with lower energy, i.e., smaller giant component. Particularly, at zero temperature $\beta\to\infty$, only the configuration with the smallest giant component exists with non-zero probability. In this way, the optimal percolation problem can be interpreted as an extreme case of the large deviations of percolation.

Recently, properties of large deviations of percolation have been analyzed using Monte Carlo Markov Chains \cite{hartmann2011large} and Belief Propagation \cite{bianconi2018rare}. In particular, Bianconi \cite{altarelli2013large,bianconi2018rare} developed a large deviation theory of percolation that characterizes the response of a sparse network to rare events. This general theory contains both continuous transitions observed for random initial damage and discontinuous transitions corresponding to rate configurations of the initial damage that suppresses the GC size. This large deviation theory of percolation was also generalized to multiplex networks \cite{bianconi2019large}, based on which a new metric, sageguard centrality, was developed to single out the nodes that control the response of the entire multiplex network to random damage \cite{coghi2018controlling}. It was found that the sageguard centrality correlates well with nodes in the optimal percolation problem.

\subsection{Summary}

It is interesting that the optimal percolation, or network dismantle problem, can be solved from quite different approaches: the CI algorithm optimizes the stability of zero solution by minimizing the spectral radius of the NB matrix; the BPD and network dismantle algorithms aim to optimally remove cycles in the network; the EI algorithm attempts to gradually identify less vital nodes so that an explosive collapse of network would occur if the remaining critical nodes are attacked; the EGP and GND algorithms work by recursively partitioning the network into equal-size components; and large deviations of percolation considers the rare events deviated from random percolation. In terms of implementation, CI proceeds as a greedy adaptive algorithm, which is straightforward to implement; the BPD, network dismantle algorithm and large deviations of percolation need to iterate BP or Min-Sum equations to find the solution; the EI algorithm iteratively selects unvaccinated nodes from a number of candidates; and the EGP and GND algorithms apply graph partition recursively with different techniques. Most of these algorithms require a reinsertion process that excludes unnecessary nodes from the optimal node set. In essence, to solve an intrinsically global optimization problem, most approaches have to transform it to another problem that can be solved locally. For instance, CI defines a centrality based on local structure; the BP equations in the BPD and network dismantle algorithms incorporate local constraints compatible with the global constraints; the score calculation in the EI algorithm depend on local connectivity; and the EGP and GND algorithms are designed to recursively partition smaller local clusters. More features of these algorithm are summarized in Table \ref{optimalpercolation}.

\begin{table}[tbhp]
\centering
\vspace{5mm}
%\resizebox{\textwidth}{!}{%
\begin{tabular}{ L{0.15\columnwidth}L{0.45\columnwidth}L{0.2\columnwidth}L{0.1\columnwidth} }
\hline\hline
Name & Description & Complexity & Ref\\
\hline
Collective Influence (CI) & Stability of the NB matrix, greedy approach, easy to interpret and implement, adapted in real-world problems & $O(N\log N)$ & \cite{morone2015influence}\\
\hline
Network dismantle & Optimal decycling, belief-propagation approach, Min-Sum algorithm, solve by iteration until convergence, compute the optimal node set simultaneously & $O(MT)$ per iteration & \cite{braunstein2016network}\\
\hline
Belief-propagation-guided decimation (BPD) & Minimum feedback vertex set, spin glass model, belief-propagation approach, no convergence needed in BP iteration, select nodes iteratively & $O(N\log N)$ & \cite{mugisha2016identifying}\\
\hline
Explosive Immunization (EI) & Explosive percolation, iteratively select less important nodes from candidates, based on score defined for each node& $O(N\log N)$ & \cite{clusella2016immunization}\\
\hline
Equal graph partitioning (EGP) & Recursively partition networks into clusters of similar size, avoid breaking small clusters & NR & \cite{chen2008finding}\\
\hline
Generalized network dismantle (GND) & Consider costs of removing nodes, node-weighted partition, recursive equal-size partition, solved by spectral properties of a Laplacian and weighted vertex cover & $O(N\log^{2+\epsilon}(N))$ & \cite{ren2019generalized}\\
\hline\hline
\end{tabular}%}
\caption{Summary of methods developed for optimal percolation. $N$ is the network size and $M$ is the number of links. ``NR'' stands for ``not reported''.}\label{optimalpercolation}
\end{table}

\section{Dynamics with continuous transitions}

The problem of influencer identification in ICMs was originated from the work of Domingos and Richardson \cite{richardson2002mining,domingos2001mining}, who aimed to advertise a product though viral marketing. Instead of viewing market as a set of independent entities, they treated it as a networked system where the potential profit contributed by a customer is mostly determined by his/her interactions with others. This problem was later formalized by Kempe {\it et al.} into a well-defined combinatorial optimization problem \cite{kempe2003maximizing}: Considering an independent cascade model in a network $G$ and an integer $k$, how to find the optimal set of $k$ seeds that initiates the largest scale propagation? The intrinsic difficulty of this problem is rooted in the exponentially growing configuration space with $k$. In fact, it was proven to be among the class of the hardest optimization problems - NP hard \cite{kempe2003maximizing}, and thus can only be solved approximately via heuristic approaches in polynomial time.

\subsection{Greedy algorithms}

One of the most intuitive solutions is to use greedy algorithm that selects the $k$ most influential single spreaders to approximate the optimal set of influencers. In this approach, the influence of single influencers can be estimated by averaging a large number of Monte Carlo simulations of spreading processes initiated by each node. As proposed in Kempe {\it et al.} \cite{kempe2003maximizing}, the optimal set of influencers $S$ is obtained by recursively adding the node that leads to the largest marginal increase to the total influence. The influence function $\sigma(S)$, defined as the expected number of active nodes given the seed set $S$, can be calculated by Monte Carlo simulations. The marginal contribution of an individual influencer $i$, $\sigma_S(i)$, can then be computed through $\sigma_S(i)=\sigma(S\cup\{i\})-\sigma(S)$. For a general class of spreading models including ICMs, the influence function $\sigma(S)$ was proven to satisfy the characteristic of the so-called submodularity \cite{cornuejols1977exceptional,nemhauser1978analysis} -- A function $\sigma(\cdot)$ is submodular if the marginal gain from adding an element to a set $S$ is at least as high as the marginal gain from adding the same element to a superset of $S$. In 1978, Nemhauser {\it et al.} mathematically proved that, for problems with submodular property, a greedy heuristic always finds a solution whose value is at least $1-[(K-1)/K]^K$ times the optimal value \cite{cornuejols1977exceptional,nemhauser1978analysis}. Here $K$ is the size of seed set. This bound has a limiting value of $1-1/e$, which is independent of the size of network or seed set. Leveraging on this theoretical result, the simple greedy algorithm for these models is guaranteed to approximate the optimal influence within a factor of $1-1/e\approx 63\%$, i.e., $\sigma(S)\geq(1-1/e)\sigma(S^*)$, where $S$ is obtained from the greedy algorithm and $S^*$ is the actual optimal set.

In case the cost of removing each node is not identical, the result of this basic greedy algorithm can be far from optimal. In such circumstance, a naive modification of the basic greedy algorithm can be made by favoring the node with maximum benefit-cost ratio. Unfortunately, this intuitive generalization can perform arbitrarily worse than the optimal solution $S^*$. In order to guarantee a relatively good performance, Leskovec {\it et al.} proposed the Cost-Effective Forward (CEF) algorithm \cite{leskovec2007cost}. As a combination of the benefit-cost and unit-cost greedy algorithms, the CEF algorithm provides a constant factor $(1-1/e)/2$ approximation of the maximal influence. Even though each of the two basic greedy algorithms can perform arbitrarily bad, it was proved that for a given circumstance, at least one of them could obtain a relatively good performance.

Due to the heavy computational burden of massive Monte Carlo simulations, greedy algorithms are unscalable to large-scale networks. This can be partly alleviated by exploiting the sparsity of cost reductions \cite{leskovec2007cost}. Furthermore, by exploiting the submodular property of the influence function, the number of simulations can be significantly reduced in practice. Given that the marginal increment of a node is monotonically decreasing with the growth of $S$, there is no need to recompute the marginal increments for all nodes at each time step. Specifically, if the marginal increment of a node $i$ in previous time steps is already smaller than that of another node $j$ in current time step, the recomputation for $\sigma(i)$ is unnecessary as it is definitely smaller than $\sigma(j)$. In calculations, the marginal influence of each node $\sigma(i)$ is marked valid initially. Before the next influencer is selected, the nodes are scanned in a decreasing order of their influence. If $\sigma(i)$ for the top node $i$ is invalid, it is recomputed and inserted into the existing order using a priority queue. If the recomputation leads to a new value that ranks at the top, it should be added into $S$ without calculating the marginal increments for any other nodes. This cost-effective lazy forward (CELF) algorithm leads to far fewer evaluations of the influence function and achieves up to a factor of 700 improvement in speed compared to CEF with equal performance. Further improvement of CELF can be made by recording the node with largest marginal gain among the nodes that are already examined in the current iteration in a heap data structure \cite{goyal2011simpath}. This technique can improve the efficiency of CELF by another 35-55\%.

Further improvement of greedy algorithms was achieved using the connection between ICMs and percolation. As indicated before, ICMs can be mapped to a bond percolation. Based on this idea, Chen {\it et al.} performed a bond percolation on a graph $G$ to estimate the influence of a seed set \cite{chen2009efficient}. Specifically, each link in a graph $G$ is randomly selected with the predefined transmission probability, and the selected links form a subgraph $G'$. Then the influence function $\sigma(S)$ can be quantified by the number of vertices reachable from $S$ in $G'$, where each edge in $G'$ is regarded as a real propagation path. With this simplification, the influence of a single node $i$ can be obtained with a linear scan of the graph $G'$ and its marginal increment to $S$ is either $0$ or $\sigma(i)$, depending on whether $i$ is in the influence range of $S$ or not. This procedure provides $O(N)$ speedup to the basic greedy algorithm. In implementation, it can be proceeded in combination with CELF to avoid unnecessary evaluations.

Despite above improvements of greedy algorithms for independent cascade model, it is still prohibitive for massively large social networks with millions of users. In order to reach the tradeoff between performance and computational efficiency, Chen {\it et al.} also proposed a heuristic degree discount algorithm \cite{chen2009efficient}. The basic idea of the degree discount algorithm is that $\sigma(i)$ should be quantified by its degree discounted by the number of its neighbors that are already included in $S$. For ICMs with a small propagation probability, the indirect influence between multi-hop neighbors is negligible so we can only take into account the direct influence between immediate neighbors. Under this assumption, a more precise metric was proposed. The performance of this algorithm nearly matches that of the basic greedy algorithms. Furthermore, it is far more efficient in combined use of the heap data structure and scalable for large-scale networks.

Another scalable variant of the basic greedy algorithm was developed based on local influence regions \cite{chen2010scalable}. The maximum influence arborescence (MIA) algorithm assumes that propagations tend to be along the maximum influence paths (MIP) between each pair of nodes, which are defined as the path with the highest propagation probability among all possible ensembles. For a given pair of nodes, the MIP between them can be computed efficiently using the Dijkstra shortest-path algorithm \cite{dijkstra1959note,cormen2009introduction}. The union of MIPs starting or ending at a node $i$ form an arborescence structure, which defines its local influence region denoted by $\delta(i)$. The global influence of a set $S$ is then quantified by the size of the union of all local influence regions: $\sigma(S)=|\bigcup_{i\in S} \delta(i)|$, where $|\cdot|$ denotes the size of a set. A tuning parameter is introduced so that all MIPs with probability below $\theta$ are discarded. By adjusting the parameter $\theta$, the size of the local influence regions can be altered so that tradeoff between computational efficiency and performance is achieved. Based on such approximations, the local marginal increment of a node can be calculated with significantly high efficiency. As the local influence function is also submodular, the basic greedy algorithm guarantees the $1-1/e$ approximation bound for influence maximization. The linearity of local marginal influence allows for the efficient update of incremental influence during iterations. More importantly, the update is only required in a local influence region around the selected influencer. 

Wang {\it et al.} proposed a community-based greedy algorithm for mining top-k influential nodes in mobile social networks \cite{wang2010community}. In the algorithm, communities with regional information diffusion are first detected, and influential nodes are then located by selecting certain communities using a dynamic programming algorithm. As shown in recent works, modularity of networks has significant impact on information diffusion \cite{nematzadeh2014optimal,curato2016optimal,yan2015global}. In the general idea, the community-based greedy algorithm considers information diffusion within each community to disentangle their interactions, thus simplifies the process of selecting multiple influencers. This algorithm was found to be more than an order of magnitude faster than typical greedy algorithms. In a recent work by Hu {\it et al.}, authors employed percolation theory to show that spreading processes of ICM are limited to a local area in most occasions \cite{hu2018local}. Therefore, local structure can identify and quantify influential global spreaders in large scale social networks. An efficient percolation-based greedy algorithm was proposed.

In another line of research, instead of using Monte Carlo simulations, centrality metrics based on the topological structure of the underlying network were adopted to estimate nodes' influence. These metrics are independent of specific spreading processes thus can be calculated with high computational efficiency. In addition, they also shed light on the impact of network topology on spreading processes, which is of great significance in both accelerating and confining propagations. Instead of actually running the spreading process, these metrics are mostly based on the local or global topology of a node in the network, for instance, number of immediate neighbors \cite{albert2000error,cohen2001breakdown,pastor2001epidemic,da2012predicting}, global position \cite{kitsak2010identification,pei2014searching,carmi2007model,zeng2013ranking,tang2015identification,malliaros2016locating}, number of shortest paths \cite{sabidussi1966centrality,freeman1977set,friedkin1991theoretical,dangalchev2006residual}, random walks \cite{brin1998anatomy,lu2011leaders,travenccolo2008accessibility}, eigenvectors \cite{bonacich1972factoring,radicchi2016leveraging,martin2014localization,restrepo2006characterizing}, path counting \cite{katz1953new,bauer2012identifying,klemm2012measure,lawyer2015understanding}, etc. Even though the optimal metric that performs best for all spreading dynamics on all underlying networks does not seem to exist \cite{gu2017ranking,borge2012absence,chami2017diffusion}, these centrality-based approaches are still persistently used due to their simplicity and relative satisfactory performance in some occasions. 

\subsection{Message passing approach}

Although the greedy optimization guarantees to approximate the maximum influence by a constant factor, it often suffers from the drawback of being trapped into local optimum. From an optimization point of view, the message-passing approach, which has been well developed in statistical physics \cite{mezard2009information,karrer2010message}, can avoid such undesirable situation. In addition, message-passing algorithms usually scales almost linearly with the number of edges, which makes it applicable to large real-world networks. Based on message-passing approach, Altarelli {\it et al.} developed the belief-propagation (BP) and max-sum (MS) algorithms for the problem of optimal immunization for SIR and SIS model \cite{altarelli2014containing}. 

For each configuration $\mathbf{s}=(s_1,s_2,...,s_N)$, the following energy function is considered
\begin{equation}
\varepsilon(\mathbf{s},\mathbf{m})=\mu\sum_{i\in V}s_ic_i+\epsilon \sum_{i\in V}m_i,
\end{equation}
where $s_i\in \{0,1\}$ ($s_i=1$ if $i$ is immunized, and $s_i=0$ otherwise), $c_i$ is the cost of immunizing node $i$ and $m_i$ is the probability that $i$ is eventually infected in the case of SIR model, or the probability that it is infected in the stationary state in the case of SIS model. The parameters $\mu$ and $\epsilon$ control the tradeoff between the cost of immunization and the cost in treating infected patients. The constraint on all feasible configurations is manifested by the local update equations of $m_i$. Based on the energy function
$\varepsilon(\mathbf{s},\mathbf{m})$, a Boltzmann weight $e^{-\beta\mathcal{E}}$ is assigned to each feasible configuration, where $\beta$ is the inverse temperature. Take the SIR model for an example, the probability $m_{ij}$ that node $i$ is infected in the absence of it neighboring node $j$ satisfies a set of equations:
\begin{equation}
m_{ij}=q+(1-q)\left[1-\prod_{k\in \partial i \setminus j}(1-pm_{ki})\right],
\end{equation}
where $q$ is the self-infection probability, $p$ is the transmission probability, and $\partial i\setminus j$ denotes the neighbors of node $i$ excluding $j$. Then the marginal probability $m_i$
that node $i$ is eventually infected is
\begin{equation}
m_i=q+(1-q)\left[1-\prod_{k\in \partial i}(1-pm_{ki})\right].
\end{equation}
Based on the locally-tree like assumption, BP equations can be derived and solved through iteration making use of the properties of convolutions of messages. As $\beta\to\infty$, the Boltzmann distribution is concentrated on the optimal configuration with the lowest energy cost. In addition, the MS equations can be developed to find the nearly optimal set of immunized nodes. In simulations, MS algorithm performs better than the topological-based heuristics, greedy algorithm as well as simulated annealing. 

In a recent work by Min \cite{min2018identifying}, the message-passing approach was used to calculate analytically the expected size of epidemic outbreaks originated from a single seed. It was found that, while the probability of triggering an epidemic outbreak depends on the location of the seed, the final size of the outbreak is insensitive to the seed once it occurs. This approach is also applicable to weighted networks.

For ICMs, two important problems are connected: the optimal selection of nodes to either minimize or maximize the influence. The minimization problem, equivalent to optimal percolation, aims to find the ``superblockers'' that should be removed to make $G_{\infty}$ as small as possible. Instead, ``superspreaders'' are those that maximize the average influence if selected as seeds. Radicchi and Castellano performed an extensive analysis over a range of real-world networks and found that these two optimization problems are not equivalent, i.e., superblockers are not superspreaders \cite{radicchi2017fundamental}. The identification of superblockers is based purely on the topology of the network, while superspreaders in influence maximization problem are strongly dependent on the parameters of the spreading dynamics. 

\subsection{Sequential seeding}

In above discussed studies, influencers or information seeds are activated simultaneously at the start of diffusion (i.e., single stage seeding). An alternative approach would be to  initiate seeds sequentially, which allows the diffusion take place before next seeds are selected. Such sequential seeding strategy has the advantage of avoiding selecting highly ranked nodes that are already activated by previous diffusion. Jankowski {\it et al.} introduced several approaches for sequential seeding, and discussed the balance between diffusion speed and coverage \cite{Jankowski2017balancing}. Using experiments in real-world networks, it was found that sequential seeding strategies achieve better coverage than single stage seeding in about 90\% of cases. Longer seeding sequences can activate more nodes but prolong the duration of diffusion. Authors proposed several variants of sequential seeding to resolve the trade-off between diffusion coverage and speed.

Jankowski {\it et al.} further presented a formal proof that sequential seeding performs at least as good as the single stage seeding does in terms of spread coverage \cite{Jankowski2018probing}. It was shown that, under mild assumptions, sequential seeding outperforms single stage seeding using the same number of seeds and node ranking. Authors compared single stage and sequential approaches with the greedy approach in experiments on directed and undirected graphs, and demonstrated that applying sequential seeding to a simple degree-based ranking leads to higher diffusion coverage than the computationally expensive greedy algorithm.

\begin{table}[tbhp]
\centering
\vspace{5mm}
%\resizebox{\textwidth}{!}{%
\begin{tabular}{ L{0.2\columnwidth}L{0.6\columnwidth}l }
\hline\hline
Name & Description & Ref\\
\hline
Monte Carlo simulations & Greedy approach, submodular function, performance guaranteed within a factor $(1-1/e)$& \cite{kempe2003maximizing}\\
\hline
Cost-Effective Forward algorithm (CEF) & Consider non-unit cost, performance guaranteed within a factor $(1-1/e)/2$& \cite{leskovec2007cost}\\
\hline
Cost-Effective lazy forward (CELF) & Fewer Monte Carlo simulations, higher efficiency, heap structure & \cite{leskovec2007cost}\\
\hline
Percolation-based approach & Map to bond percolation, use subgraph to estimate influence &\cite{chen2009efficient}\\
\hline
Degree discount algorithm  & Direct influence between immediate neighbors & \cite{chen2009efficient}\\
\hline
Maximum influence arborescence (MIA) & Maximum influence path, Dijkstra shortest-path algorithm, arborescence structure, tradeoff between computational efficiency and performance & \cite{chen2010scalable}\\
\hline
Community-based algorithm & Community detection, dynamic programming algorithm & \cite{wang2010community}\\
\hline
Message-passing approach & Belief propagation, Max-Sum algorithm, SIR and SIS model, solved through iteration &\cite{altarelli2014containing}\\
\hline
Message-passing approach & Expected size of epidemic outbreaks, insensitive to origin, applicable to weighted networks & \cite{min2018identifying}\\
\hline
Sequential seeding & Initiate information seeds sequentially, trade-off between coverage and speed &\cite{Jankowski2017balancing}\\
\hline\hline
\end{tabular}%}
\caption{Summary of some methods developed for influence maximization in ICMs.}\label{continuous}
\end{table}

\subsection{Summary}

We summarize features of the methods introduced in this section in Table \ref{continuous}. For greedy approaches, the central task is to estimate the influence of each node, using either Monte Carlo simulation or local structural information. Following this idea, its improvement is designed along two directions: avoiding unnecessary simulations or develop better local proxies for influence. The performance of greedy algorithms is guaranteed for dynamics with submodular property. The message-passing approach calculates the spreading outcomes by solving a set of BP equations, thus considers the problem from a global viewpoint. In addition, there is no requirement for the submodular property. The sequential seeding strategy aims to maximize diffusion coverage by adopting an alternative seeding approach, which brings our attention to the trade-off between diffusion coverage and speed.

In a recent work by Erkol {\it et al.} \cite{erkol2019systematic}, the performance of 16 methods for identifying influential spreaders in ICMs were systematically compared on a large corpus of 100 real-world networks. Extensive numerical experiments indicate that the performance of many simple heuristic methods, such as adaptive degree and closeness centrality, is similar to that of more computationally expensive greedy algorithms. This provides some practical methods for large-scale problems where greedy algorithms are prohibitive. It was also found that the performance can be further improved towards the optimality by using hybrid methods that combine multiple topological metrics.

\section{Dynamics with discontinuous transitions}

Threshold models and k-core percolation are frequently used to describe cascading processes with discontinuous phase transitions in various disciplines, for instance, failure propagation in infrastructure \cite{buldyrev2010catastrophic}, diffusion of innovations in social networks \cite{rogers2010diffusion}, and adoption of new behaviors \cite{centola2010spread}. By definition, k-core percolation is a special case of a more general class of threshold models where each node has a fixed threshold $k$. The fundamental difference from threshold models to ICMs is that, in threshold models, the state of a node is collectively determined by the states of all its neighbors. As a consequence, the impact of perturbing one node can propagate to a vast area of the network through long-range chains of interactions, manifested by a discontinuous phase transition in network dynamics. In this section, we first introduce methods developed for linear threshold models (LTMs) using greedy strategy, belief-propagation and collective influence, and then discuss algorithms designed for k-core percolation. Note that algorithms designed for LTMs are applicable to k-core percolation.

\subsection{Linear threshold models}

Linear threshold models have several different forms. A typical LTM is defined on a weighted network $G=(V,E,\omega)$, where $\omega: V\times V \rightarrow [0, 1]$ is a weight function and $\omega =0$ iff the corresponding edge does not exist. Similar to ICMs, the spreading process in LTMs is initiated by a set of seeds while all other nodes are inactive. In following steps, a node is activated if the sum of weights of its active neighbors reaches its predefined threshold value $\theta_i$, i.e. $\sum_{j\in\partial i}\omega_{ij}\geq\theta_i$, where $\partial i$ stands for the set of neighbors of node $i$. In another form, a node is activated if it has at least a certain number of active neighbors. 

The threshold value for each node can be either a fixed constant or a random variable drawn from a predefined distribution. For LTMs with a uniform fixed threshold value $\omega\in[0,1]$, Singh {\it et al.} studied the cascade size as a function of the fraction of seeds \cite{singh2013threshold}. It was found that even for large threshold values, a critical fraction of seeds exists beyond which the cascade becomes global. In addition, networks with community structure and high clustering were found more effective in facilitating cascade than homogeneous random networks. For LTMs with heterogeneous thresholds, Karampourniotis {\it et al.} examined how cascade size varies with the standard deviation of the distribution of thresholds \cite{karampourniotis2015impact}. Using a truncated normal distribution, authors varied the distribution of thresholds between two extreme cases: identical thresholds and a uniform distribution. A non-monotonic change in the cascade size appeared with the varying standard deviation, indicating that, for a given number of seeds, an optimal variance of the threshold distribution exists.

\subsubsection{Greedy approach}

The greedy algorithm is also applicable to LTMs. For a special class of LTMs where the weight of each edge and the threshold of each node are drawn uniformly from the interval $[0,1]$, it was proved that its influence function is submodular \cite{kempe2003maximizing}. Therefore, the influence maximization problem in this class of LTMs can be approximately solved by greedy algorithms. 

Like ICMs, a linear threshold model can be also mapped to a modified percolation process defined as follows: Each node $i$ picks at most one of its incoming edges, with probability $\omega_{ji}$ to select the edge from $j$ to $i$ and $1-\sum_j \omega_{ji}$ to select none. The selected edges are defined as live. Considering the subgraph $G'$ composed of live edges, Kempe {\it et al.} proved that for a given set $S$, the number of nodes activated by $S$ in LTMs has the same distribution with the number of reachable nodes of $S$ in the subgraph $G'$ \cite{kempe2003maximizing}.

Using the same mapping, Chen {\it et al.} gave an efficient approximation of the influence of an individual node in a local subgraph \cite{chen2010scalable}. In cases where the weights $\omega_{ij}$ and $\omega_{ji}$ are not symmetrical, the undirected graph $G$ can be transformed into an equivalent directed graph, where edges from $i$ to $j$ and from $j$ to $i$ are both included. Using the randomized algorithm of Cohen \cite{cohen1997size}, the influence of a set $S$ is quantified by the number of nodes reachable from $S$ in the subgraph $G'$. Although computing the exact influence in a network is $\#$P-hard, this approximation based on directed acyclic graph (DAG) can be finished within linear time. In order to further accelerate the calculation, a local DAG (LDAG) is considered instead of DAG. Validation of this approximation is supported by the exponential decay of influence with the propagation length. The construction of LDAG should include a majority part of influence from other nodes while discarding the nodes with small influence. Similar to the idea in Ref. \cite{chen2010scalable}, a threshold is introduced to control the size of LDAG, so that the tradeoff between efficiency and accuracy can be tuned. Once the LDAG is constructed, the incremental influence of each node can be quantified with great efficiency. As a result, the LDAG algorithm is scalable to networks with millions of nodes and is among the best greedy algorithms in performance.

The LDAG algorithm assumes that the influence of a node is mainly bounded within its LDAG. However, if the spreading process starting from a node can reach outside its LDAG, the estimation of influence in the LDAG algorithm might be inaccurate. Besides, the algorithm depends heavily on the proper choice of a high quality LDAG, which is an NP-hard problem itself. To avoid these problems, Goyal {\it et al.} developed the SIMPATH algorithm in which the influence of a node is quantified by enumerating the simple paths starting from it \cite{goyal2011simpath}. Although this problem is also $\#$P-hard, it can be well approximated with high efficiency by enumerating paths within a small neighborhood. With this approximation, the influence of a set $S$ can be calculated as the sum of influence of each node in it on appropriately induced subgraphs. Similar to the arborescence structures constructed in Ref. \cite{chen2010scalable}, a tuning parameter is introduced to control the size of the neighborhood, which leads to a direct trade-off between the accuracy and computational efficiency. To reduce the number of estimation calls in SIMPATH, a vertex cover optimization was introduced so that only the influence of nodes in the vertex cover set needs to be computed. For the rest of the nodes, their influence can be derived from their neighbors. Besides, as the seed set $S$ grows larger, a look ahead optimization can be made to accelerate the estimation: It picks the top $l$ most promising candidates as a batch in the start of an iteration and shares the marginal gain computation within the batch. Extensive experiments on real datasets show that compared with the basic greedy algorithm, the SIMPATH algorithm is more efficient, consumes less memory and produces seed sets with larger influence.

In a recent study by Karampourniotis {\it et al.}, two different metrics were proposed to find influencers for LTMs with fixed heterogeneous thresholds \cite{Karampourniotis2019influence}. The first metric, termed Balanced Index (BI), tends to select nodes with high resistance to activation and those with large out-degree. BI is a linear combination of three properties of a node including degree, susceptibility to new information, and the impact its activation would have on its neighbors. The performance of BI depends on the weights of these three properties. The second metric, termed Group Performance Index (GPI), quantifies the impact of each node as a seed when it is part of randomly selected seed set. For LTMs with fixed and known thresholds, these two metrics were found effective for influence maximization.

The performance of most greedy algorithms mentioned above is guaranteed thanks to the submodular property of the influence function. However, for a general LTM with fixed weights and thresholds, the influence function is not always submodular \cite{kempe2003maximizing}. An important class of LTM that may not be submodular is defined as follows: A node $i$ is activated only after a certain number $m_i$ of its neighbors are activated. The variation of threshold $m_i$ can lead to two qualitatively different classes of cascades featured by either continuous or discontinuous phase transitions. For instance, in the special case when $m_i=k_i-1$ ($k_i$ is the degree of node $i$), the scale of propagation experiences a continuous phase transition \cite{morone2015influence}. In contrast, for k-core percolation and bootstrap percolation, a first-order, or discontinuous phase transition may appear \cite{goltsev2006k}. Solutions to the influence maximization problem in LTMs without submodular property require a better understanding of the physical mechanism of the spreading process, and will be introduced in detail in following subsections.

\subsubsection{Belief-propagation algorithms}

For the influence maximization problem on a general LTM, Altarelli {\it et al.} regarded it as a nontypical trajectory deviated from the average behavior of dynamics initiated by randomly chosen seeds \cite{altarelli2013large}. To explore the dynamical properties of nontypical trajectories of general LTMs, Altarelli {\it et al.} proposed a BP algorithm that could estimate statistical properties of nontypical trajectories and found the initial conditions that lead to cascading with desired properties \cite{altarelli2013optimizing}. In contrast to ICMs, the trajectory of a given LTM is determined solely by its initial condition. Due to the irreversibility of LTM dynamics, the spreading process can be parameterized by a configuration $\mathbf{t}=(t_1,t_2,...t_N)$, where $t_i\in \mathbf{T}=\{0,1,...T,\infty\}$ is the activation time of node $i$. Considering the properties of LTM, the dynamical rule can be represented by the constraint on the activation time of a node and its neighbors \cite{altarelli2013optimizing}: $t_i=\phi_i(\{t_j\})$, where
\begin{equation}
t_i=\phi_i(\{t_j\})=\min\left\{t\in \mathbf{T}:\sum_{j\in \partial i} \omega_{ji}\mathbb{I}[t_j<t]\geq \theta_i\right\}.
\end{equation}
Based on this static parametrization of LTM, the following Boltzmann distribution is considered:
\begin{equation}
P(\mathbf{t})=\frac{1}{Z}e^{-\beta\varepsilon(\mathbf{t})}\prod_{i\in V}\psi_i(t_i,\{t_j\}),
\end{equation}
where $\psi_i(t_i,\{t_j\})=\mathbb{I}[t_i=0]+\mathbb{I}[t_i=\phi_i(\{t_j\})]$, $Z=\sum_{\mathbf{t}}e^{-\beta\varepsilon(\mathbf{t})}\prod_{i\in V}\psi_i(t_i,\{t_j\})$. The most common form of the energy function is $\varepsilon(\mathbf{t})=\sum_i\varepsilon_i(t_i)$, where $\varepsilon_i(t_i)=\mathbb{I}[t_i=0]-\varepsilon\mathbb{I}[t_i<\infty]$. For $\varepsilon=0$, the distribution degenerates to the spreading dynamics initiated by a random set of seeds.

In order to avoid short loops in the factor graph that describes the constraints of a configuration, a dual factor graph is constructed with a variable node $(t_i,t_j)$ introduced to each edge $(i,j)$. The obtained dual factor graph is locally tree-like if the original network is so. This allows for the application of the cavity method. Denote $P_j(t_j)$ as the marginal probability that node $j$ is activated at time $t_j$. In a tree-like factor graph, it can be calculated as
\begin{equation}
P_j(t_j)\propto\sum_{\{t_i\}_{i\in \partial j}}e^{-\beta\varepsilon_j(t_j)}\psi_j(t_j,\{t_i\})\prod_{i\in \partial j}H_{ij}(t_i,t_j),
\end{equation}
where $H_{ij}(t_i,t_j)$ is defined as the probability that nodes $i$ and $j$ get activated at $t_i$ and $t_j$ respectively in the absence of the constraint $\psi_j$ and the energy term $\varepsilon_j$. This equation computes the contribution from all neighbors of node $j$. In the dual factor graph, $H_{ij}(t_i,t_j)$, named cavity marginals or ``beliefs'', satisfy local constraints described by a set of belief-propagation (BP) equations. In particular, the recursive relation of the cavity marginal $H_{ij}(t_i,t_j)$ on the dual factor graph defines the following belief BP equations \cite{altarelli2013optimizing}:
\begin{equation}
H_{ij}(t_i,t_j)\propto e^{-\beta\varepsilon_i(t_i)}\sum_{\{t_k\}}\psi_i(t_i,\{t_k\})\prod_kH_{ki}(t_k,t_i).
\end{equation}
Here $\psi_i(t_i,\{t_k\})$ is the local constraint on links connected to node $i$ (except node $j$), the product term computes the contribution of ``beliefs'' from the neighbors of node $i$ excluding node $j$, the summation term considers different occasions of $t_k$ for neighbors of node $i$, and $e^{-\beta\varepsilon_i(t_i)}$ defines the weight for energy $\varepsilon_i(t_i)$ using the Boltzmann distribution. The BP equations are solved through iteration. Once the fixed values of the cavity marginals are obtained, the marginal $P_j(t_j)$ and other statistics of nontypical trajectories, such as the entropy and distribution of activation time, can be subsequently computed.

On homogeneous random regular graphs, the BP equations can be simplified to a self-consistent equation of a single marginal. Analysis for different threshold values indicates quantitative difference in the distribution of activation time $P(t)$ for the regimes of continuous and discontinuous transitions. Specifically, for continuous transitions, $P(t)$ is monotonically decreasing. On the contrary, $P(t)$ shows a second peak for discontinuous transitions, corresponding to the abrupt cascade activation. In order to obtain the optimal set of seeds, Max-Sum equations can be derived by setting the inverse temperature $\beta\to\infty$ in the energy function \cite{altarelli2013optimizing}. Authors performed numerical experiments on a real-world network (the Epinions network) with an energy function $\varepsilon(\mathbf{t})=\sum_i\{c_i\mathbb{I}[t_i=0]-r_i\mathbb{I}[t_i<\infty]\}$, where $c_i$ is the cost of seeding node $i$ and $r_i$ is the revenue generated by the activation of node $r$. The Max-Sum algorithm was compared with competing methods including greedy algorithm based on energy computation (GA), greedy algorithm based on HITS (HITS), high degree (Hubs) and simulated annealing (SA). The Max-Sum algorithm outperforms other approaches by selecting the seed set that best tradeoffs the revenue and cost. The performance of Max-Sum algorithms on synthetic networks also outperforms a range of centrality metrics, as shown in Fig. \ref{CITM}.

Extending the work under the assumption of replica symmetry, Guggiola and Semerjian \cite{guggiola2015minimal} studied the minimal contagious set problem for LTM dynamics with and without a constraint on the maximal activation time $T$. In this theoretically impressive work, authors aim to find the theoretical limit of the minimal contagious set (i.e., the minimal seed set that can activate the entire graph) in random regular graphs using the cavity method with the effect of replica symmetry breaking. Following the theoretical development, a survey propagation like algorithm \cite{mezard2002random} is investigated on single instances of random regular graphs to find the exact seed set. It was found that the survey propagation algorithm achieves near-optimal performance for small activation time limit. For a large activation time limit, authors reported convergence issues in iteration that cannot be effectively solved by a simple damping. However, stopping the iterations after a predefined time proved to be a pragmatic and satisfactory strategy. In this work, authors tested the algorithm on random regular graphs; in practice, how survey propagation algorithm works for more realistic networks needs to be tested. Readers interested in the survey propagation algorithm can find details in Ref. \cite{guggiola2015minimal}.

\subsubsection{Collective influence in threshold model}

The collective influence theory can be generalized to deal with the influence maximization in general LTMs \cite{pei2017efficient}. For a network $G(V,E)$ with $N$ nodes and $M$ links, we use the vector $\mathbf{n}=(n_1, n_2, \cdots, n_N)$ to record whether a node $i$ is chosen as a seed ($n_i=1$) or not ($n_i=0$). The LTM spreading starts from a $q = \sum_i n_i/N$ fraction of active seeds and evolves following a threshold rule: a node $i$ becomes active if it has at least $m_i$ active neighbors. Here, the threshold $m_i$ is an integer ranging from 1 to the degree of node $i$. Further, we introduce $\nu_i$ to indicate the final state of node $i$: active ($\nu_i=1$) or inactive ($\nu_i=0$). For a given $q$ fraction of seeds, the influence maximization problem aims to find the optimal set of seeds so that the size of active population is maximized.

\begin{figure}[]
  \centering
  \includegraphics[width=1\columnwidth]{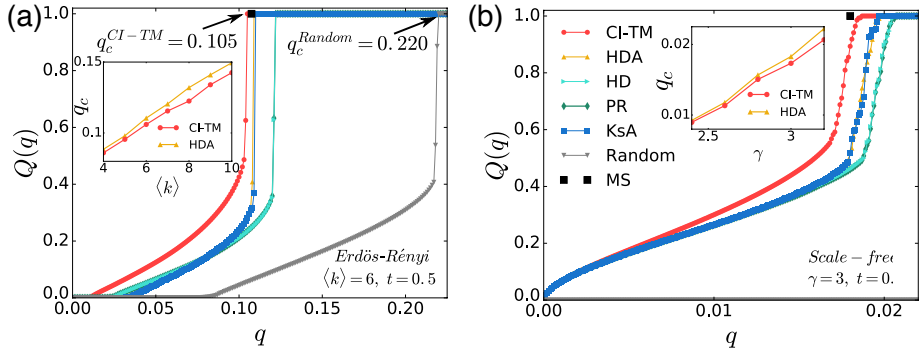}
  \caption{{Performance of the CI-TM algorithm in ER and SF networks.}
   (a), Size of active giant component $Q(q)$ as a function of the fraction of seed $q$ for ER networks ($N=2\times 10^5$, $\langle k\rangle=6$). The CI-TM algorithm is compared with high degree adaptive (HDA), high degree (HD), PageRank (PR), k-core adaptive (KsA), Random and the Max-Sum (MS) algorithm. Inset shows the critical values $q_c$ identified by HDA and CI-TM for different mean degrees. (b), Results for scale-free networks ($N=2\times 10^5$, $\gamma=3$). Inset presents the critical values $q_c$ for different power-law exponents $\gamma$. Figure reuse from \cite{pei2017efficient} is permitted by Springer Nature.} \label{CITM}
     \end{figure}

For each link $i\to j$, we introduce a binary variable $\nu_{i\to j}$ as the indicator of $i$ being in the active state assuming node $j$ is disconnected from the network. For locally tree-like networks, $\nu_{i\to j}$ satisfies a set of self-consistent message-passing equations. Different from the case of optimal percolation in Ref. \cite{morone2015influence}, the zero solution is not a fixed point. As a consequence, the stability analysis around zero solution in Ref. \cite{morone2015influence} is no longer valid for LTMs. However, the solution can be approximated through iteration of the linearized system. By linearizing the equations, it was found that the subsequent activation of nodes in each iteration only depends on the number of subcritical nodes, defined as the nodes with $m_i-1$ active neighbors (i.e., nodes whose activation can be triggered by one more active neighbor). Moreover, subcritical nodes can form long subcritical paths that generate long-range cascade of activation, which is core to the discontinuous transition in LTM dynamics. Following this idea, the CI-TM (Collective Influence in Threshold Model) algorithm was proposed that recursively selects nodes with the largest CI-TM score. The CI-TM score enumerates the number of subcritical paths starting from each node, and uses that to quantify nodes' spreading capability. With an $O(N\log N)$ computational complexity, the efficient CI-TM algorithm is applicable to large-scale networks. In numerical simulations, the CI-TM algorithm outperforms the greedy algorithm and several widely used heuristic centralities, and achieves comparable performance to the Max-Sum algorithm in synthetic random networks (see Fig. \ref{CITM}).

\subsection{K-core percolation}

Because k-core percolation is a special case of LTMs, influence maximization algorithms developed for general LTMs can be naturally extended to work for k-core percolation. 

In statistical physics and combinatorial optimization, several theoretical works have explored the lower and upper bounds on the size of the minimal set to destroy the k-core. In the evaluation of approximating algorithms, these results can help us to assess how far the estimated size of minimal contagious set is from the theoretical limit. For instance, Bau {\it et al.} studied the decycling numbers of random regular graphs \cite{bau2002decycling}. As stated before, the decycling process is equivalent to destroying the 2-core of networks. For a random cubic graph $G$ that all nodes have degree 3, it was proven that the decycling number $\phi(G)=\lceil N/4+1/2\rceil$ as the graph size $N\to \infty$. For a general random $d$-regular graph $G$ with $N$ nodes ($d\geq4$), authors proved that $\phi(G)/N$ is bounded below and above asymptotically almost surely by certain constants that depend solely on $d$. In particular, the lower and upper bounds can be calculated by solving an algebraic equation and a set of differential equations, respectively. Janson and Thomason found that, for sparse random graphs or random regular networks with $N$ nodes with $N\to\infty$, the number of nodes that must be removed so that no component with more than $k$ nodes exists is essentially the same for all values of $k$ if $k\to\infty$ and $k=o(N)$ \cite{janson2008dismantling}. Reichman showed that the size of a contagious set is bounded from above by $\sum_{v\in V}\min\left\{1,\frac{k}{d(v)+1}\right\}$ in the destruction of k-core ($d(v)$ is the degree of node $v$) \cite{reichman2012new}. Later, using the cavity method with replica symmetry breaking, Guggiola and Semerjian \cite{guggiola2015minimal} obtained several conjectures on the size of minimal contagious sets for k-core percolation in random regular graphs. In particular, authors conjectured that the minimal contagious set size is 1/6 for 5-regular random graphs with a threshold of 3, and 1/4 for 6-regular with threshold 4. In addition, they also proposed the conjecture for ($k$+1)-regular networks with the threshold $k$ that the minimal contagious set size is $1-2(\ln k)/k-2/k+O(1/(k\ln k))$. According to this conjecture, the minimal contagious set size 3-regular (cubic) random graphs with a threshold of 2 is 1/4, which is in agreement with the decycling number of cubic random graphs $\phi(G)/N\to 1/4$ obtained in Ref. \cite{bau2002decycling}. Sun {\it et al.} also proposed a lower bound of the network dismantling problem by analyzing specific 2-core subnetworks of many real-world networks that have heterogeneous degree distribution \cite{sun2018lower}. Coja-Oghlan {\it et al.} explored the minimal contagious set problem on graphs with expansion properties \cite{coja2015contagious}.

Recently, Schmidt {\it et al.} \cite{schmidt2019minimal} studied the minimal contagious sets for k-core percolation in random networks. In this work, authors proposed a generalized CoreHD algorithm, in which nodes with the highest degree in the k-core are recursively removed until the k-core completely collapses. To analyze the property of this algorithm, the generalized CoreHD-guided k-core removal was translated to a random process on the degree distribution of the graph \cite{wormald1995differential,wormald1999differential}. The running time of the process, characterized by a set of nonlinear ordinary differential equations, describes the behavior of the algorithm on a random graph. By analyzing the stopping time, new upper bounds on the minimal contagious set were obtained, which improve the best currently known ones in Ref. \cite{guggiola2015minimal,bau2002decycling}. This approach is applicable not only to random regular graphs, but also to random networks generated from the configuration model with a given degree distribution. Inspired by the analysis of the CoreHD algorithm, an improved algorithm, called WEAK-NEIGHBOR, was developed. In this algorithm, instead of removing high degree nodes, nodes with the highest value $k_i-\sum_{j\in\partial i}k_j/k_i$ in the k-core are removed ($k_i$ is the degree of node $i$). For networks with bounded degree, the algorithm has $O(N)$ complexity, where $N$ is the network size. In numerical experiments, the WEAK-NEIGHBOR algorithm improves over the generalized CoreHD algorithm and CI-TM algorithm in a range of k-core percolation processes in random regular graphs.

\subsection{Summary}

For LTMs, the major effort in greedy approach is to develop more efficient and accurate estimation of marginal increments using local network structure. This pursuit has inspired different techniques designed for this goal. Most greedy methods quantify the marginal increment by the number of nodes that would be activated if a node is selected as a seed. The CI-TM algorithm, in contrast, uses the number of subcritical paths attached to a node to estimate the marginal increment. Belief-propagation approaches solve the problem as a global issue through iteration, and can flexibly incorporate the cost of activating seeds. Apart from devising practical methods to solve the influence maximization problem for LTMs, analytical works on random regular graphs would help to identify how far away current approaches are from the theoretical limit of the size of optimal seed set. Features about the introduced methods are summarized in Table \ref{discontinuous}.

\begin{table}[tbhp]
\centering
\vspace{5mm}
%\resizebox{\textwidth}{!}{%
\begin{tabular}{ L{0.2\columnwidth}L{0.6\columnwidth}l}
\hline\hline
Name & Description & Ref\\
\hline
Monte Carlo simulations & Greedy approach, submodular function, performance guaranteed within a factor $(1-1/e)$ & \cite{kempe2003maximizing}\\
\hline
Percolation-based approach & Map to bond percolation, use subgraph to estimate influence, $O(N)$ complexity &\cite{chen2010scalable}\\
\hline
Local directed acyclic graph (LDAG) & Decay of influence with the propagation length, discard the nodes with small influence &\cite{chen2010scalable}\\
\hline
SIMPATH algorithm & Enumerating simple paths, look ahead optimization, trade-off between accuracy and efficiency & \cite{goyal2011simpath}\\
\hline
Balanced Index (BI) & Select nodes with high resistance to activation and with large out-degree, fast to compute & \cite{Karampourniotis2019influence}\\
\hline
Group Performance Index (GPI) & Measure performance of each node as a seed when it is a part of randomly selected seed set & \cite{Karampourniotis2019influence}\\
\hline
Belief-propagation algorithm & Large deviations of LTM dynamics, consider cost and revenue, Max-Sum equations, solved by iteration until convergence &\cite{altarelli2013optimizing}\\
\hline
Survey propagation like algorithm & Near-optimal performance, solved by iteration until convergence, applied to random regular graphs &\cite{guggiola2015minimal}\\
\hline
Collective Influence in Threshold Model (CI-TM) & Greedy approach, based on linearized message-passing equations, use subcritical clusters to estimate influence, $O(N\log N)$ complexity & \cite{pei2017efficient}\\
\hline
CoreHD & Remove high-degree nodes in 2-core, perform well on loopy networks, $O(N)$ complexity &\cite{zdeborova2016fast}\\
\hline
WEAK-NEIGHBOR & Improvement over the generalized CoreHD algorithm and CI-TM algorithm, $O(N)$ complexity & \cite{schmidt2019minimal} \\
\hline\hline
\end{tabular}%}
\caption{Summary of some methods developed for influence maximization in LTMs and k-core percolation.}\label{discontinuous}
\end{table}

\section{Conclusions and discussions}

With an increasing number of real-world complex systems formulated as networks, a theory for identifying influencers is required to facilitate a better understanding and control of various dynamical complex systems. Over the years, this problem has been extensively studied in different contexts by physicists, mathematicians, sociologists, computer scientists, etc. In this survey, we review recent advances in this area. Because this topic spans a wide spectrum of research, we cannot report every relevant work exhaustively. However, we try to organize the survey in a way such that recent developments made in several fields of broad interest are covered.

Despite great advances in influencer identification, many ongoing problems and directions exist that need to be addressed in future works. First, as shown in several theoretical works, even for homogeneous structure such as random regular networks, there is still a gap between the result obtained from the state-of-the-art algorithms and its theoretical limit. This provides a room that we can improve in algorithm design. Second, the topological structure of real-world complex systems can be much more complicated than the case considered in ideal conditions. In a recent comparative analysis, it was found that recently proposed techniques perform well only on specific network types \cite{wandelt2018comparative}. Further, connections may be time-varying in temporal networks \cite{holme2012temporal}, or posses complicated interlayer interactions in multiplex networks \cite{boccaletti2014structure,kivela2014multilayer}. Third, in many systems, links are often of different types with distinct functions. These systems cannot be described by the simple network structure discussed before, and do not even admit a formal definition of influencers. In future works, these open problems remain to be explored in more detail.

In terms of applications, use of influencer identification theory in biological, social and engineering systems is still very limited. As some advanced methodologies in statistical physics are technical and challenging to interpret, applying the latest progresses of influencer identification in specific real-world systems can better illustrate and disseminate these techniques. Moreover, current methods are mostly developed under ideal conditions. In real-world systems, errors or noises inevitably exist \cite{orsini2015quantifying,erkol2018identifying}. How to quantify and alleviate the impact of errors or noises is of great practical values in applications. In addition, certain non-dynamical factors beyond the simplified assumption in pure modeling studies, e.g., human activity \cite{iribarren2009impact,gallos2012people,muchnik2013origins,min2015finding,aral2012identifying,aral2011identifying,aral2018social,pei2015exploring,teng2014individual}, homophily \cite{aral2009distinguishing,centola2011experimental,aral2013engineering}, complex contagion \cite{centola2010spread,centola2007complex} and social influence bias \cite{muchnik2013social}, may need to be considered. This calls for a deeper understanding of the systems under study and a more integrative application of the influencer identification theory.

\section*{Funding}
 Part of this work was supported by the National Institutes of Health [R01EB022720, U54CA137788, and U54CA132378 to H.A.M.]; National Science Foundation [1515022 to H.A.M.]; Army Research Laboratory [W911NF-09-2-0053 to H.A.M.]; and China Scholarship Council and the Academic Excellence Foundation of BUAA for PhD Students (to J.W.).

% can use a bibliography generated by BibTeX as a .bbl file
% BibTeX documentation can be easily obtained at:
% http://www.ctan.org/tex-archive/biblio/bibtex/contrib/doc/

%\bibliographystyle{imaiai}
%\bibliographystyle{unsrt}
%\bibliography{ref}

\begin{thebibliography}{99}

\bibitem{pastor2015epidemic}
\textsc{Pastor-Satorras, R., Castellano, C., Van~Mieghem, P.  {\small \&}
  Vespignani, A.}  (2015) Epidemic processes in complex networks. \emph{Rev.
  Mod. Phys.}, \textbf{87}(3), 925.

\bibitem{zhang2016dynamics}
\textsc{Zhang, Z.-K., Liu, C., Zhan, X.-X., Lu, X., Zhang, C.-X.  {\small \&}
  Zhang, Y.-C.}  (2016) Dynamics of information diffusion and its applications
  on complex networks. \emph{Phys. Rep.}, \textbf{651}, 1--34.

\bibitem{bullmore2009complex}
\textsc{Bullmore, E.  {\small \&} Sporns, O.}  (2009) Complex brain networks:
  graph theoretical analysis of structural and functional systems. \emph{Nat.
  Rev. Neurosci.}, \textbf{10}(3), 186.

\bibitem{montoya2006ecological}
\textsc{Montoya, J.~M., Pimm, S.~L.  {\small \&} Sol{\'e}, R.~V.}  (2006)
  Ecological networks and their fragility. \emph{Nature}, \textbf{442}(7100),
  259.

\bibitem{newman2003structure}
\textsc{Newman, M. E.~J.}   (2003) The structure and function of complex networks.
  \emph{SIAM Rev.}, \textbf{45}(2), 167--256.

\bibitem{barrat2008dynamical}
\textsc{Barrat, A., Barthelemy, M.  {\small \&} Vespignani, A.}  (2008)
  \emph{Dynamical processes on complex networks}. Cambridge University Press,
  New York.

\bibitem{boccaletti2006complex}
\textsc{Boccaletti, S., Latora, V., Moreno, Y., Chavez, M.  {\small \&} Hwang,
  D.-U.}  (2006) Complex networks: Structure and dynamics. \emph{Phys. Rep.},
  \textbf{424}(4-5), 175--308.

\bibitem{albert2002statistical}
\textsc{Albert, R.  {\small \&} Barab{\'a}si, A.-L.}  (2002) Statistical
  mechanics of complex networks. \emph{Rev. Mod. Phys.}, \textbf{74}(1), 47.

\bibitem{watts2007influentials}
\textsc{Watts, D.~J.  {\small \&} Dodds, P.~S.}  (2007) Influentials, networks,
  and public opinion formation. \emph{J. Consumer. Res.}, \textbf{34}(4),
  441--458.

\bibitem{del2018finding}
\textsc{Del~Ferraro, G., Moreno, A., Min, B., Morone, F.,
  P{\'e}rez-Ram{\'\i}rez, {\'U}., P{\'e}rez-Cervera, L., Parra, L.~C., Holodny,
  A., Canals, S.  {\small \&} Makse, H.~A.}  (2018) Finding influential nodes
  for integration in brain networks using optimal percolation theory.
  \emph{Nat. Comm.}, \textbf{9}(1), 2274.

\bibitem{reis2014avoiding}
\textsc{Reis, S.~D., Hu, Y., Babino, A., Andrade~Jr, J.~S., Canals, S., Sigman,
  M.  {\small \&} Makse, H.~A.}  (2014) Avoiding catastrophic failure in
  correlated networks of networks. \emph{Nat. Phys.}, \textbf{10}(10), 762.

\bibitem{zamora2010cortical}
\textsc{Zamora-L{\'o}pez, G., Zhou, C.  {\small \&} Kurths, J.}  (2010)
  Cortical hubs form a module for multisensory integration on top of the
  hierarchy of cortical networks. \emph{Front. Neuroinform.}, \textbf{4}, 1.

\bibitem{may1972will}
\textsc{May, R.~M.}  (1972) Will a large complex system be stable?.
  \emph{Nature}, \textbf{238}(5364), 413.

\bibitem{scheffer2012anticipating}
\textsc{Scheffer, M., Carpenter, S.~R., Lenton, T.~M., Bascompte, J., Brock,
  W., Dakos, V., Van~de Koppel, J., Van~de Leemput, I.~A., Levin, S.~A.,
  Van~Nes, E.~H.  et~al.}  (2012) Anticipating critical transitions.
  \emph{Science}, \textbf{338}(6105), 344--348.

\bibitem{mills1993keystone}
\textsc{Mills, L.~S., Soul{\'e}, M.~E.  {\small \&} Doak, D.~F.}  (1993) The
  keystone-species concept in ecology and conservation. \emph{BioScience},
  \textbf{43}(4), 219--224.

\bibitem{morone2018kcore}
\textsc{Morone, F., Del~Ferraro, G.  {\small \&} Makse, H.~A.}  (2019) The
  k-core as a predictor of structural collapse in mutualistic ecosystems.
  \emph{Nat. Phys.}, \textbf{15}(1), 95.

\bibitem{kitsak2010identification}
\textsc{Kitsak, M., Gallos, L.~K., Havlin, S., Liljeros, F., Muchnik, L.,
  Stanley, H.~E.  {\small \&} Makse, H.~A.}  (2010) Identification of
  influential spreaders in complex networks. \emph{Nat. Phys.}, \textbf{6}(11),
  888.

\bibitem{Pei2019inference}
\textsc{Pei, S., Morone, F., Liljeros, F., Makse, H.~A.  {\small \&} Shaman,
  J.}  (2018) Inference and control of the nosocomial transmission of
  Methicillin-resistant Staphylococcus aureus. \emph{eLife}, \textbf{7},
  e40977.

\bibitem{freeman1978centrality}
\textsc{Freeman, L.~C.}  (1978) Centrality in social networks conceptual
  clarification. \emph{Social Networks}, \textbf{1}(3), 215--239.

\bibitem{morone2015influence}
\textsc{Morone, F.  {\small \&} Makse, H.~A.}  (2015) Influence maximization in
  complex networks through optimal percolation. \emph{Nature},
  \textbf{524}(7563), 65.

\bibitem{braunstein2016network}
\textsc{Braunstein, A., Dall’Asta, L., Semerjian, G.  {\small \&}
  Zdeborov{\'a}, L.}  (2016) Network dismantling. \emph{Proc. Natl. Acad. Sci.
  USA}, \textbf{113}(44), 12368--12373.

\bibitem{kempe2003maximizing}
\textsc{Kempe, D., Kleinberg, J.  {\small \&} Tardos, {\'E}.}  (2003)
  Maximizing the spread of influence through a social network. in
  \emph{Proceedings of the 9th ACM SIGKDD International Conference on Knowledge
  Discovery and Data Mining}, pp. 137--146. ACM.

\bibitem{leskovec2007dynamics}
\textsc{Leskovec, J., Adamic, L.~A.  {\small \&} Huberman, B.~A.}  (2007) The
  dynamics of viral marketing. \emph{ACM Transactions on the Web (TWEB)},
  \textbf{1}(1), 5.

\bibitem{richardson2002mining}
\textsc{Richardson, M.  {\small \&} Domingos, P.}  (2002) Mining
  knowledge-sharing sites for viral marketing. in \emph{Proceedings of the 8th
  ACM SIGKDD International Conference on Knowledge Discovery and Data Mining},
  pp. 61--70. ACM.

\bibitem{pastor2002immunization}
\textsc{Pastor-Satorras, R.  {\small \&} Vespignani, A.} (2002) Immunization of complex networks. \emph{Phys. Rev.
  E}, \textbf{65}(3), 036104.

\bibitem{chen2008finding}
\textsc{Chen, Y., Paul, G., Havlin, S., Liljeros, F.  {\small \&} Stanley,
  H.~E.}  (2008) Finding a better immunization strategy. \emph{Phys. Rev.
  Lett.}, \textbf{101}(5), 058701.

\bibitem{cohen2003efficient}
\textsc{Cohen, R., Havlin, S.  {\small \&} Ben-Avraham, D.}  (2003) Efficient
  immunization strategies for computer networks and populations. \emph{Phys.
  Rev. Lett.}, \textbf{91}(24), 247901.

\bibitem{albert2000error}
\textsc{Albert, R., Jeong, H.  {\small \&} Barab{\'a}si, A.-L.}  (2000) Error
  and attack tolerance of complex networks. \emph{Nature}, \textbf{406}(6794),
  378.

\bibitem{cohen2001breakdown}
\textsc{Cohen, R., Erez, K., Ben-Avraham, D.  {\small \&} Havlin, S.}  (2001)
  Breakdown of the Internet under intentional attack. \emph{Phys. Rev. Lett.},
  \textbf{86}(16), 3682.

\bibitem{latora2005vulnerability}
\textsc{Latora, V.  {\small \&} Marchiori, M.}  (2005) Vulnerability and
  protection of infrastructure networks. \emph{Phys. Rev. E}, \textbf{71}(1),
  015103.

\bibitem{keeling2011modeling}
\textsc{Keeling, M.~J.  {\small \&} Rohani, P.}  (2011) \emph{Modeling
  infectious diseases in humans and animals}. Princeton University Press,
  Princeton.

\bibitem{thebault2010stability}
\textsc{Th{\'e}bault, E.  {\small \&} Fontaine, C.}  (2010) Stability of
  ecological communities and the architecture of mutualistic and trophic
  networks. \emph{Science}, \textbf{329}(5993), 853--856.

\bibitem{kandel2000principles}
\textsc{Kandel, E.~R., Schwartz, J.~H., Jessell, T.~M., of~Biochemistry, D.,
  Jessell, M. B.~T., Siegelbaum, S.  {\small \&} Hudspeth, A.}  (2000)
  \emph{Principles of neural science}, vol.~4. McGraw-Hill, New York.

\bibitem{davidson2005gene}
\textsc{Davidson, E.  {\small \&} Levin, M.}  (2005) Gene regulatory networks.
  \emph{Proc. Natl. Acad. Sci. USA}, \textbf{102}(14), 4935--4935.

\bibitem{erdos1960evolution}
\textsc{Erd\H{o}s, P.  {\small \&} R{\'e}nyi, A.}  (1960) On the evolution of
  random graphs. \emph{Publ. Math. Inst. Hung. Acad. Sci.}, \textbf{5}(1),
  17--60.

\bibitem{callaway2000network}
\textsc{Callaway, D.~S., Newman, M. E.~J., Strogatz, S.~H.  {\small \&} Watts,
  D.~J.}  (2000) Network robustness and fragility: Percolation on random
  graphs. \emph{Phys. Rev. Lett.}, \textbf{85}(25), 5468.

\bibitem{newman2002spread}
\textsc{Newman, M. E.~J.}  (2002) Spread of epidemic disease on networks.
  \emph{Phys. Rev. E}, \textbf{66}(1), 016128.

\bibitem{seidman1983network}
\textsc{Seidman, S.~B.}  (1983) Network structure and minimum degree.
  \emph{Social Networks}, \textbf{5}(3), 269--287.

\bibitem{dorogovtsev2006k}
\textsc{Dorogovtsev, S.~N., Goltsev, A.~V.  {\small \&} Mendes, J. F.~F.}
  (2006) K-core organization of complex networks. \emph{Phys. Rev. Lett.},
  \textbf{96}(4), 040601.

\bibitem{kermack1932contributions}
\textsc{Kermack, W.~O.  {\small \&} McKendrick, A.~G.}  (1932) Contributions to
  the mathematical theory of epidemics. II.—The problem of endemicity.
  \emph{Proc. R. Soc. Lond. A}, \textbf{138}(834), 55--83.

\bibitem{dorogovtsev2008critical}
\textsc{Dorogovtsev, S.~N., Goltsev, A.~V.  {\small \&} Mendes, J.~F.}  (2008)
  Critical phenomena in complex networks. \emph{Rev. Mod. Phys.},
  \textbf{80}(4), 1275.

\bibitem{baxter2010bootstrap}
\textsc{Baxter, G.~J., Dorogovtsev, S.~N., Goltsev, A.~V.  {\small \&} Mendes,
  J.~F.}  (2010) Bootstrap percolation on complex networks. \emph{Phys. Rev.
  E}, \textbf{82}(1), 011103.

\bibitem{granovetter1978threshold}
\textsc{Granovetter, M.}  (1978) Threshold models of collective behavior.
  \emph{Am. J. Sociol.}, \textbf{83}(6), 1420--1443.

\bibitem{goltsev2006k}
\textsc{Goltsev, A.~V., Dorogovtsev, S.~N.  {\small \&} Mendes, J. F.~F.}
  (2006) k-core (bootstrap) percolation on complex networks: Critical phenomena
  and nonlocal effects. \emph{Phys. Rev. E}, \textbf{73}(5), 056101.

\bibitem{watts2002simple}
\textsc{Watts, D.~J.}  (2002) A simple model of global cascades on random
  networks. \emph{Proc. Natl. Acad. Sci. USA}, \textbf{99}(9), 5766--5771.

\bibitem{altarelli2014containing}
\textsc{Altarelli, F., Braunstein, A., Dall’Asta, L., Wakeling, J.~R.
  {\small \&} Zecchina, R.}  (2014) Containing epidemic outbreaks by
  message-passing techniques. \emph{Phys. Rev. X}, \textbf{4}(2), 021024.

\bibitem{altarelli2013optimizing}
\textsc{Altarelli, F., Braunstein, A., Dall’Asta, L.  {\small \&} Zecchina,
  R.} (2013) Optimizing spread dynamics on graphs by message
  passing. \emph{J. Stat. Mech.}, \textbf{2013}(09), P09011.

\bibitem{mugisha2016identifying}
\textsc{Mugisha, S.  {\small \&} Zhou, H.-J.}  (2016) Identifying optimal
  targets of network attack by belief propagation. \emph{Phys. Rev. E},
  \textbf{94}(1), 012305.

\bibitem{clusella2016immunization}
\textsc{Clusella, P., Grassberger, P., P{\'e}rez-Reche, F.~J.  {\small \&}
  Politi, A.}  (2016) Immunization and targeted destruction of networks using
  explosive percolation. \emph{Phys. Rev. Lett.}, \textbf{117}(20), 208301.

\bibitem{pei2013spreading}
\textsc{Pei, S.  {\small \&} Makse, H.~A.}  (2013) Spreading dynamics in
  complex networks. \emph{J. Stat. Mech.}, \textbf{2013}(12), P12002.

\bibitem{lu2016vital}
\textsc{L{\"u}, L., Chen, D., Ren, X.-L., Zhang, Q.-M., Zhang, Y.-C.  {\small
  \&} Zhou, T.}  (2016) Vital nodes identification in complex networks.
  \emph{Phys. Rep.}, \textbf{650}, 1--63.

\bibitem{pei2018theories}
\textsc{Pei, S., Morone, F.  {\small \&} Makse, H.~A.}  (2018) Theories for
  influencer identification in complex networks. in \emph{Complex Spreading
  Phenomena in Social Systems}, pp. 125--148. Springer, Cham, Switzerland.

\bibitem{bollobas1998random}
\textsc{Bollob{\'a}s, B.}  (1998) Random graphs. in \emph{Modern graph theory},
  pp. 215--252. Springer, New York.

\bibitem{newman2001random}
\textsc{Newman, M. E.~J., Strogatz, S.~H.  {\small \&} Watts, D.~J.}  (2001)
  Random graphs with arbitrary degree distributions and their applications.
  \emph{Phys. Rev. E}, \textbf{64}(2), 026118.

\bibitem{barabasi1999emergence}
\textsc{Barab{\'a}si, A.-L.  {\small \&} Albert, R.}  (1999) Emergence of
  scaling in random networks. \emph{Science}, \textbf{286}(5439), 509--512.

\bibitem{clauset2009power}
\textsc{Clauset, A., Shalizi, C.~R.  {\small \&} Newman, M. E.~J.}  (2009)
  Power-law distributions in empirical data. \emph{SIAM Rev.}, \textbf{51}(4),
  661--703.

\bibitem{radicchi2015predicting}
\textsc{Radicchi, F.}  (2015) Predicting percolation thresholds in networks.
  \emph{Phys. Rev. E}, \textbf{91}(1), 010801.

\bibitem{karrer2014percolation}
\textsc{Karrer, B., Newman, M. E.~J.  {\small \&} Zdeborov{\'a}, L.}  (2014)
  Percolation on sparse networks. \emph{Phys. Rev. Lett.}, \textbf{113}(20),
  208702.

\bibitem{pastor2001epidemic}
\textsc{Pastor-Satorras, R.  {\small \&} Vespignani, A.} (2001) Epidemic spreading in scale-free networks.
  \emph{Phys. Rev. Lett.}, \textbf{86}(14), 3200.

\bibitem{altarelli2013large}
\textsc{Altarelli, F., Braunstein, A., Dall’Asta, L.  {\small \&} Zecchina,
  R.}  (2013) Large deviations of cascade processes on graphs. \emph{Phys.
  Rev. E}, \textbf{87}(6), 062115.

\bibitem{hartmann2011large}
\textsc{Hartmann, A.~K.}  (2011) Large-deviation properties of largest
  component for random graphs. \emph{Eur. Phys. J. B}, \textbf{84}(4),
  627--634.

\bibitem{bianconi2018rare}
\textsc{Bianconi, G.}  (2018) Rare events and discontinuous percolation
  transitions. \emph{Phys. Rev. E}, \textbf{97}(2), 022314.

\bibitem{bianconi2019large}
\textsc{Bianconi, G.}  (2019) Large deviation theory of percolation on multiplex
  networks. \emph{J. Stat. Mech.}, \textbf{2019}(2), 023405.

\bibitem{coghi2018controlling}
\textsc{Coghi, F., Radicchi, F.  {\small \&} Bianconi, G.}  (2018) Controlling
  the uncertain response of real multiplex networks to random damage.
  \emph{Phys. Rev. E}, \textbf{98}(6), 062317.

\bibitem{hethcote2000mathematics}
\textsc{Hethcote, H.~W.}  (2000) The mathematics of infectious diseases.
  \emph{SIAM Rev.}, \textbf{42}(4), 599--653.

\bibitem{anderson1992infectious}
\textsc{Anderson, R.~M.  {\small \&} May, R.~M.}  (1992) \emph{Infectious
  diseases of humans: dynamics and control}. Oxford University Press, Oxford.

\bibitem{diekmann2000mathematical}
\textsc{Diekmann, O.  {\small \&} Heesterbeek, J. A.~P.}  (2000)
  \emph{Mathematical epidemiology of infectious diseases: model building,
  analysis and interpretation}, vol.~5. John Wiley \& Sons, West Sussex,
  England.

\bibitem{pastor2001epidemicpre}
\textsc{Pastor-Satorras, R.  {\small \&} Vespignani, A.}  (2001) Epidemic
  dynamics and endemic states in complex networks. \emph{Phys. Rev. E},
  \textbf{63}(6), 066117.

\bibitem{moreno2004dynamics}
\textsc{Moreno, Y., Nekovee, M.  {\small \&} Pacheco, A.~F.}  (2004) Dynamics
  of rumor spreading in complex networks. \emph{Phys. Rev. E}, \textbf{69}(6),
  066130.

\bibitem{li2014rumor}
\textsc{Li, W., Tang, S., Pei, S., Yan, S., Jiang, S., Teng, X.  {\small \&}
  Zheng, Z.}  (2014) The rumor diffusion process with emerging independent
  spreaders in complex networks. \emph{Physica A}, \textbf{397}, 121--128.

\bibitem{yan2014dynamical}
\textsc{Yan, S., Tang, S., Pei, S., Jiang, S.  {\small \&} Zheng, Z.}  (2014)
  Dynamical immunization strategy for seasonal epidemics. \emph{Phys. Rev. E},
  \textbf{90}(2), 022808.

\bibitem{altarelli2014bayesian}
\textsc{Altarelli, F., Braunstein, A., Dall’Asta, L., Lage-Castellanos, A.
  {\small \&} Zecchina, R.}  (2014) Bayesian inference of epidemics on networks
  via belief propagation. \emph{Phys. Rev. Lett.}, \textbf{112}(11), 118701.

\bibitem{lokhov2014inferring}
\textsc{Lokhov, A.~Y., M{\'e}zard, M., Ohta, H.  {\small \&} Zdeborov{\'a}, L.}
   (2014) Inferring the origin of an epidemic with a dynamic message-passing
  algorithm. \emph{Phys. Rev. E}, \textbf{90}(1), 012801.

\bibitem{shah2010detecting}
\textsc{Shah, D.  {\small \&} Zaman, T.}  (2010) Detecting sources of computer
  viruses in networks: theory and experiment. in \emph{ACM SIGMETRICS
  Performance Evaluation Review}, pp. 203--214. ACM.

 
\bibitem{pei2015detecting}
\textsc{Pei, S., Tang, S.  {\small \&} Zheng, Z.}  (2015) Detecting the influence of spreading in social networks with excitable sensor networks. \emph{PLoS One}, \textbf{10}(5), e0124848.

\bibitem{shah2011rumors}
\textsc{Shah, D.  {\small \&} Zaman, T.}  (2011) Rumors in a network: Who's the culprit?. \emph{IEEE
  Trans. Inf. Theory}, \textbf{57}(8), 5163--5181.

\bibitem{comin2011identifying}
\textsc{Comin, C.~H.  {\small \&} da~Fontoura~Costa, L.}  (2011) Identifying
  the starting point of a spreading process in complex networks. \emph{Phys.
  Rev. E}, \textbf{84}(5), 056105.

\bibitem{pei2018forecasting}
\textsc{Pei, S., Kandula, S., Yang, W.  {\small \&} Shaman, J.}  (2018)
  Forecasting the spatial transmission of influenza in the United States.
  \emph{Proc. Natl. Acad. Sci. USA}, \textbf{115}(11), 2752--2757.
  
\bibitem{scarpino2019predictability}
\textsc{Scarpino, S.~V.  {\small \&} Petri, G.}  (2019) On the predictability
  of infectious disease outbreaks. \emph{Nat. Comm.}, \textbf{10}(1), 898.

\bibitem{pei2017counteracting}
\textsc{Pei, S.  {\small \&} Shaman, J.}  (2017) Counteracting structural
  errors in ensemble forecast of influenza outbreaks. \emph{Nat. Comm.},
  \textbf{8}(1), 925.

\bibitem{kandula2018evaluation}
\textsc{Kandula, S., Yamana, T., Pei, S., Yang, W., Morita, H.  {\small \&}
  Shaman, J.}  (2018) Evaluation of mechanistic and statistical methods in
  forecasting influenza-like illness. \emph{J. Royal Soc. Interface},
  \textbf{15}(144), 20180174.
  
\bibitem{pei2019predictability}
\textsc{Pei, S., Cane, M.~A. {\small \&} Shaman, J.}  (2019) Predictability in 
process-based ensemble forecast of influenza. \emph{PLoS Comput. Biol.},
  \textbf{15}(2), e1006783.


\bibitem{batagelj2011fast}
\textsc{Batagelj, V.  {\small \&} Zaver{\v{s}}nik, M.}  (2011) Fast algorithms
  for determining (generalized) core groups in social networks. \emph{Advances
  in Data Analysis and Classification}, \textbf{5}(2), 129--145.

\bibitem{lu2016h}
\textsc{L{\"u}, L., Zhou, T., Zhang, Q.-M.  {\small \&} Stanley, H.~E.}  (2016)
  The H-index of a network node and its relation to degree and coreness.
  \emph{Nat. Comm.}, \textbf{7}, 10168.

\bibitem{hirsch2005index}
\textsc{Hirsch, J.~E.}  (2005) An index to quantify an individual's scientific
  research output. \emph{Proc. Natl. Acad. Sci. USA}, \textbf{102}(46),
  16569--16572.

\bibitem{azimi2014k}
\textsc{Azimi-Tafreshi, N., G{\'o}mez-Gardenes, J.  {\small \&} Dorogovtsev,
  S.}  (2014) k- core percolation on multiplex networks. \emph{Phys. Rev. E},
  \textbf{90}(3), 032816.

\bibitem{chalupa1979bootstrap}
\textsc{Chalupa, J., Leath, P.~L.  {\small \&} Reich, G.~R.}  (1979) Bootstrap
  percolation on a Bethe lattice. \emph{J. Phys. C}, \textbf{12}(1), L31.

\bibitem{cellai2011tricritical}
\textsc{Cellai, D., Lawlor, A., Dawson, K.~A.  {\small \&} Gleeson, J.~P.}
  (2011) Tricritical point in heterogeneous k-core percolation. \emph{Phys.
  Rev. Lett.}, \textbf{107}(17), 175703.

\bibitem{baxter2011heterogeneous}
\textsc{Baxter, G.~J., Dorogovtsev, S.~N., Goltsev, A.~V.  {\small \&} Mendes,
  J.~F.}  (2011) Heterogeneous k-core versus bootstrap percolation on
  complex networks. \emph{Phys. Rev. E}, \textbf{83}(5), 051134.

\bibitem{cellai2013critical}
\textsc{Cellai, D., Lawlor, A., Dawson, K.~A.  {\small \&} Gleeson, J.~P.}  (2013) Critical phenomena in heterogeneous k-core
  percolation. \emph{Phys. Rev. E}, \textbf{87}(2), 022134.

\bibitem{dodds2004universal}
\textsc{Dodds, P.~S.  {\small \&} Watts, D.~J.}  (2004) Universal behavior in a
  generalized model of contagion. \emph{Phys. Rev. Lett.}, \textbf{92}(21),
  218701.
  
\bibitem{morone2019jamming}  
\textsc{Morone, F., Burleson-Lesser, K., Vinutha, H.~A., Sastry, S. {\small \&} Makse, H.~A.} (2019) The jamming transition is a k-core percolation transition. \emph{Physica A}, \textbf{516}, 172-177.


\bibitem{azimi2019generalization}
\textsc{Azimi-Tafreshi, N., Osat, S.  {\small \&} Dorogovtsev, S.}  (2019)
  Generalization of core percolation on complex networks. \emph{Phys. Rev. E},
  \textbf{99}(2), 022312.

\bibitem{buldyrev2010catastrophic}
\textsc{Buldyrev, S.~V., Parshani, R., Paul, G., Stanley, H.~E.  {\small \&}
  Havlin, S.}  (2010) Catastrophic cascade of failures in interdependent
  networks. \emph{Nature}, \textbf{464}(7291), 1025.

\bibitem{yang2017small}
\textsc{Yang, Y., Nishikawa, T.  {\small \&} Motter, A.~E.}  (2017) Small
  vulnerable sets determine large network cascades in power grids.
  \emph{Science}, \textbf{358}(6365), eaan3184.

\bibitem{dorogovtsev2006k-core}
\textsc{Dorogovtsev, S., Goltsev, A.  {\small \&} Mendes, J.}  (2006) k-core
  architecture and k-core percolation on complex networks. \emph{Physica D:
  Nonlinear Phenomena}, \textbf{224}(1-2), 7--19.

\bibitem{schwarz2006onset}
\textsc{Schwarz, J., Liu, A.~J.  {\small \&} Chayes, L.}  (2006) The onset of
  jamming as the sudden emergence of an infinite k-core cluster.
  \emph{Europhys. Lett.}, \textbf{73}(4), 560.

\bibitem{parshani2010interdependent}
\textsc{Parshani, R., Buldyrev, S.~V.  {\small \&} Havlin, S.}  (2010)
  Interdependent networks: Reducing the coupling strength leads to a change
  from a first to second order percolation transition. \emph{Phys. Rev. Lett.},
  \textbf{105}(4), 048701.

\bibitem{gao2011robustness}
\textsc{Gao, J., Buldyrev, S.~V., Havlin, S.  {\small \&} Stanley, H.~E.}
  (2011) Robustness of a network of networks. \emph{Phys. Rev. Lett.},
  \textbf{107}(19), 195701.

\bibitem{guggiola2015minimal}
\textsc{Guggiola, A.  {\small \&} Semerjian, G.}  (2015) Minimal contagious
  sets in random regular graphs. \emph{J. Stat. Phys.}, \textbf{158}(2),
  300--358.

\bibitem{ackerman2010combinatorial}
\textsc{Ackerman, E., Ben-Zwi, O.  {\small \&} Wolfovitz, G.}  (2010)
  Combinatorial model and bounds for target set selection. \emph{Theor. Comput.
  Sci.}, \textbf{411}(44-46), 4017--4022.

\bibitem{dreyer2009irreversible}
\textsc{Dreyer~Jr, P.~A.  {\small \&} Roberts, F.~S.}  (2009) Irreversible
  k-threshold processes: Graph-theoretical threshold models of the spread of
  disease and of opinion. \emph{Discrete Appl. Math.}, \textbf{157}(7),
  1615--1627.

\bibitem{reichman2012new}
\textsc{Reichman, D.}  (2012) New bounds for contagious sets. \emph{Discrete
  Math.}, \textbf{312}(10), 1812--1814.

\bibitem{feige2017contagious}
\textsc{Feige, U., Krivelevich, M.  {\small \&} Reichman, D.}  (2017)
  Contagious sets in random graphs. \emph{Ann. Appl. Probab.}, \textbf{27}(5),
  2675--2697.

\bibitem{angel2017minimal}
\textsc{Angel, O.  {\small \&} Kolesnik, B.}  (2017) Large deviations for subcritical bootstrap percolation on the random graph.
  \emph{arXiv preprint arXiv:1705.06815}.

\bibitem{angel2016thresholds}
\textsc{Angel, O.  {\small \&} Kolesnik, B.}  (2016) Thresholds for contagious
  sets in random graphs. \emph{arXiv preprint arXiv:1611.10167}.

\bibitem{hashimoto1989zeta}
\textsc{Hashimoto, K.-i.}  (1989) Zeta functions of finite graphs and
  representations of p-adic groups. \emph{Adv. Stud. Pure Math.}, \textbf{15},
  211--280.

\bibitem{saad2011numerical}
\textsc{Saad, Y.}  (2011) \emph{Numerical methods for large eigenvalue
  problems: revised edition}, vol.~66. SIAM, Philadelphia.

\bibitem{morone2016collective}
\textsc{Morone, F., Min, B., Bo, L., Mari, R.  {\small \&} Makse, H.~A.}
  (2016) Collective influence algorithm to find influencers via optimal
  percolation in massively large social media. \emph{Sci. Rep.}, \textbf{6},
  30062.

\bibitem{bau2002decycling}
\textsc{Bau, S., Wormald, N.~C.  {\small \&} Zhou, S.}  (2002) Decycling
  numbers of random regular graphs. \emph{Random Struct. Alg.},
  \textbf{21}(3-4), 397--413.

\bibitem{kobayashi2016fragmenting}
\textsc{Kobayashi, T.  {\small \&} Masuda, N.}  (2016) Fragmenting networks by
  targeting collective influencers at a mesoscopic level. \emph{Sci. Rep.},
  \textbf{6}, 37778.

\bibitem{osat2017optimal}
\textsc{Osat, S., Faqeeh, A.  {\small \&} Radicchi, F.}  (2017) Optimal
  percolation on multiplex networks. \emph{Nat. Comm.}, \textbf{8}(1), 1540.

\bibitem{teng2016collective}
\textsc{Teng, X., Pei, S., Morone, F.  {\small \&} Makse, H.~A.}  (2016)
  Collective influence of multiple spreaders evaluated by tracing real
  information flow in large-scale social networks. \emph{Sci. Rep.},
  \textbf{6}, 36043.

\bibitem{bovet2018validation}
\textsc{Bovet, A., Morone, F.  {\small \&} Makse, H.~A.}  (2018) Validation of
  Twitter opinion trends with national polling aggregates: Hillary Clinton vs
  Donald Trump. \emph{Sci. Rep.}, \textbf{8}(1), 8673.

\bibitem{bovet2019influence}
\textsc{Bovet, A.  {\small \&} Makse, H.~A.}  (2019) Influence of fake news in
  Twitter during the 2016 US presidential election. \emph{Nat. Comm.},
  \textbf{10}(1), 7.

\bibitem{morone2017model}
\textsc{Morone, F., Roth, K., Min, B., Stanley, H.~E.  {\small \&} Makse,
  H.~A.}  (2017) Model of brain activation predicts the neural collective
  influence map of the brain. \emph{Proc. Natl. Acad. Sci. USA},
  \textbf{114}(15), 3849--3854.

\bibitem{luo2017inferring}
\textsc{Luo, S., Morone, F., Sarraute, C., Travizano, M.  {\small \&} Makse,
  H.~A.}  (2017) Inferring personal economic status from social network
  location. \emph{Nat. Comm.}, \textbf{8}, 15227.

\bibitem{szolnoki2016collective}
\textsc{Szolnoki, A.  {\small \&} Perc, M.}  (2016) Collective influence in
  evolutionary social dilemmas. \emph{EPL}, \textbf{113}(5), 58004.

\bibitem{zhang2018dynamic}
\textsc{Zhang, R.  {\small \&} Pei, S.}  (2018) Dynamic range maximization in
  excitable networks. \emph{Chaos}, \textbf{28}(1), 013103.

\bibitem{wang2018optimal}
\textsc{Wang, J., Pei, S., Wei, W., Feng, X.  {\small \&} Zheng, Z.}  (2018)
  Optimal stabilization of Boolean networks through collective influence.
  \emph{Phys. Rev. E}, \textbf{97}(3), 032305.

\bibitem{wang2019on}
\textsc{Wang, J., Zhang, R., Wei, W., Pei, S.  {\small \&} Zheng, Z.}  (2019)
  On the stability of multilayer Boolean networks under targeted immunization.
  \emph{Chaos}, \textbf{29}(1), 013133.
  
 \bibitem{pei2012how}
\textsc{Pei, S., Tang, S., Yan, S., Jiang, S., Zhang, X. {\small \&} Zheng, Z.} (2012) 
How to enhance the dynamic range of excitatory-inhibitory excitable networks. 
\emph{Phys. Rev. E}, \textbf{86}(2), 021909. 
  

\bibitem{karp1972reducibility}
\textsc{Karp, R.~M.}  (1972) Reducibility among combinatorial problems. in
  \emph{Complexity of Computer Computations}, pp. 85--103. Springer, New York.

\bibitem{marinari2004circuits}
\textsc{Marinari, E.  {\small \&} Monasson, R.}  (2004) Circuits in random
  graphs: from local trees to global loops. \emph{J. Stat. Mech.},
  \textbf{2004}(09), P09004.

\bibitem{marinari2005algorithm}
\textsc{Marinari, E., Monasson, R.  {\small \&} Semerjian, G.}  (2005) An
  algorithm for counting circuits: Application to real-world and random graphs.
  \emph{EPL}, \textbf{73}(1), 8.

\bibitem{bianconi2005loops}
\textsc{Bianconi, G.  {\small \&} Marsili, M.}  (2005) Loops of any size and
  Hamilton cycles in random scale-free networks. \emph{J. Stat. Mech.},
  \textbf{2005}(06), P06005.

\bibitem{zhou2013spin}
\textsc{Zhou, H.-J.}  (2013) Spin glass approach to the feedback vertex set
  problem. \emph{Eur. Phys. J. B}, \textbf{86}(11), 455.

\bibitem{im2018dismantling}
\textsc{Im, Y.~S.  {\small \&} Kahng, B.}  (2018) Dismantling Efficiency and
  Network Fractality. \emph{Phys. Rev. E}, \textbf{98}(1), 012316.

\bibitem{mezard2009information}
\textsc{Mezard, M.  {\small \&} Montanari, A.}  (2009) \emph{Information,
  Physics, and Computation}. Oxford University Press, Oxford.

\bibitem{mezard2001bethe}
\textsc{M{\'e}zard, M.  {\small \&} Parisi, G.}  (2001) The Bethe lattice spin
  glass revisited. \emph{Eur. Phys. J. B}, \textbf{20}(2), 217--233.

\bibitem{bayati2008statistical}
\textsc{Bayati, M., Borgs, C., Braunstein, A., Chayes, J., Ramezanpour, A.
  {\small \&} Zecchina, R.}  (2008) Statistical mechanics of steiner trees.
  \emph{Phys. Rev. Lett.}, \textbf{101}(3), 037208.

\bibitem{zdeborova2016fast}
\textsc{Zdeborov{\'a}, L., Zhang, P.  {\small \&} Zhou, H.-J.}  (2016) Fast and
  simple decycling and dismantling of networks. \emph{Sci. Rep.}, \textbf{6},
  37954.

\bibitem{schmidt2019minimal}
\textsc{Schmidt, C., Pfister, H.~D.  {\small \&} Zdeborov{\'a}, L.}  (2019)
  Minimal sets to destroy the k-core in random networks. \emph{Phys. Rev. E},
  \textbf{99}(2), 022310.

\bibitem{achlioptas2009explosive}
\textsc{Achlioptas, D., D'souza, R.~M.  {\small \&} Spencer, J.}  (2009)
  Explosive percolation in random networks. \emph{Science}, \textbf{323}(5920),
  1453--1455.

\bibitem{da2010explosive}
\textsc{da~Costa, R.~A., Dorogovtsev, S.~N., Goltsev, A.~V.  {\small \&}
  Mendes, J. F.~F.}  (2010) Explosive percolation transition is actually
  continuous. \emph{Phys. Rev. Lett.}, \textbf{105}(25), 255701.

\bibitem{riordan2011explosive}
\textsc{Riordan, O.  {\small \&} Warnke, L.}  (2011) Explosive percolation is
  continuous. \emph{Science}, \textbf{333}(6040), 322--324.

\bibitem{grassberger2011explosive}
\textsc{Grassberger, P., Christensen, C., Bizhani, G., Son, S.-W.  {\small \&}
  Paczuski, M.}  (2011) Explosive percolation is continuous, but with unusual
  finite size behavior. \emph{Phys. Rev. Lett.}, \textbf{106}(22), 225701.

\bibitem{friedman2009construction}
\textsc{Friedman, E.~J.  {\small \&} Landsberg, A.~S.}  (2009) Construction and
  analysis of random networks with explosive percolation. \emph{Phys. Rev.
  Lett.}, \textbf{103}(25), 255701.

\bibitem{newman2001fast}
\textsc{Newman, M. E.~J.  {\small \&} Ziff, R.~M.}  (2001) Fast Monte Carlo
  algorithm for site or bond percolation. \emph{Phys. Rev. E}, \textbf{64}(1),
  016706.

\bibitem{lipton1979generalized}
\textsc{Lipton, R.~J., Rose, D.~J.  {\small \&} Tarjan, R.~E.}  (1979)
  Generalized nested dissection. \emph{SIAM J. Numer. Anal.}, \textbf{16}(2),
  346--358.

\bibitem{ren2019generalized}
\textsc{Ren, X.-L., Gleinig, N., Helbing, D.  {\small \&} Antulov-Fantulin, N.}
   (2019) Generalized network dismantling. \emph{Proc. Natl. Acad. Sci. USA},
  \textbf{116}(14), 6554--6559.

\bibitem{bar1981linear}
\textsc{Bar-Yehuda, R.  {\small \&} Even, S.}  (1981) A linear-time
  approximation algorithm for the weighted vertex cover problem. \emph{Journal
  of Algorithms}, \textbf{2}(2), 198--203.

\bibitem{domingos2001mining}
\textsc{Domingos, P.  {\small \&} Richardson, M.}  (2001) Mining the network
  value of customers. in \emph{Proceedings of the 7th ACM SIGKDD International
  Conference on Knowledge Discovery and Data Mining}, pp. 57--66. ACM.

\bibitem{cornuejols1977exceptional}
\textsc{Cornuejols, G., Fisher, M.~L.  {\small \&} Nemhauser, G.~L.}  (1977)
  Location of bank accounts to optimize float: An analytic study of exact and
  approximate algorithms. \emph{Manag. Sci.}, \textbf{23}(8), 789--810.

\bibitem{nemhauser1978analysis}
\textsc{Nemhauser, G.~L., Wolsey, L.~A.  {\small \&} Fisher, M.~L.}  (1978) An
  analysis of approximations for maximizing submodular set functions—I.
  \emph{Math. Program.}, \textbf{14}(1), 265--294.

\bibitem{leskovec2007cost}
\textsc{Leskovec, J., Krause, A., Guestrin, C., Faloutsos, C., VanBriesen, J.
  {\small \&} Glance, N.}  (2007) Cost-effective outbreak detection in
  networks. in \emph{Proceedings of the 13th ACM SIGKDD International
  Conference on Knowledge Discovery and Data Mining}, pp. 420--429. ACM.

\bibitem{goyal2011simpath}
\textsc{Goyal, A., Lu, W.  {\small \&} Lakshmanan, L.~V.}  (2011) SIMPATH: An
  efficient algorithm for influence maximization under the linear threshold
  model. in \emph{Data Mining (ICDM), 2011 IEEE 11th International Conference
  on}, pp. 211--220. IEEE.

\bibitem{chen2009efficient}
\textsc{Chen, W., Wang, Y.  {\small \&} Yang, S.}  (2009) Efficient influence
  maximization in social networks. in \emph{Proceedings of the 15th ACM SIGKDD
  International Conference on Knowledge Discovery and Data Mining}, pp.
  199--208. ACM.

\bibitem{chen2010scalable}
\textsc{Chen, W., Wang, C.  {\small \&} Wang, Y.}  (2010) Scalable influence
  maximization for prevalent viral marketing in large-scale social networks. in
  \emph{Proceedings of the 16th ACM SIGKDD International Conference on
  Knowledge Discovery and Data Mining}, pp. 1029--1038. ACM.

\bibitem{dijkstra1959note}
\textsc{Dijkstra, E.~W.}  (1959) A note on two problems in connexion with
  graphs. \emph{Numerische Mathematik}, \textbf{1}(1), 269--271.

\bibitem{cormen2009introduction}
\textsc{Cormen, T.~H., Leiserson, C.~E., Rivest, R.~L.  {\small \&} Stein, C.}
  (2009) \emph{Introduction to algorithms}. MIT Press, Cambridge.

\bibitem{wang2010community}
\textsc{Wang, Y., Cong, G., Song, G.  {\small \&} Xie, K.}  (2010)
  Community-based greedy algorithm for mining top-k influential nodes in mobile
  social networks. in \emph{Proceedings of the 16th ACM SIGKDD International
  Conference on Knowledge Discovery and Data Mining}, pp. 1039--1048. ACM.

\bibitem{nematzadeh2014optimal}
\textsc{Nematzadeh, A., Ferrara, E., Flammini, A.  {\small \&} Ahn, Y.-Y.}
  (2014) Optimal network modularity for information diffusion. \emph{Phys. Rev.
  Lett.}, \textbf{113}(8), 088701.

\bibitem{curato2016optimal}
\textsc{Curato, G.  {\small \&} Lillo, F.}  (2016) Optimal information
  diffusion in stochastic block models. \emph{Phys. Rev. E}, \textbf{94}(3),
  032310.

\bibitem{yan2015global}
\textsc{Yan, S., Tang, S., Fang, W., Pei, S.  {\small \&} Zheng, Z.}  (2015)
  Global and local targeted immunization in networks with community structure.
  \emph{J. Stat. Mech.}, \textbf{2015}(8), P08010.

\bibitem{hu2018local}
\textsc{Hu, Y., Ji, S., Jin, Y., Feng, L., Stanley, H.~E.  {\small \&} Havlin,
  S.}  (2018) Local structure can identify and quantify influential global
  spreaders in large scale social networks. \emph{Proc. Natl. Acad. Sci. USA},
  \textbf{115}(29), 7468--7472.

\bibitem{da2012predicting}
\textsc{Da~Silva, R. A.~P., Viana, M.~P.  {\small \&} da~Fontoura~Costa, L.}
  (2012) Predicting epidemic outbreak from individual features of the
  spreaders. \emph{J. Stat. Mech.}, \textbf{2012}(07), P07005.

\bibitem{pei2014searching}
\textsc{Pei, S., Muchnik, L., Andrade~Jr, J.~S., Zheng, Z.  {\small \&} Makse,
  H.~A.}  (2014) Searching for superspreaders of information in real-world
  social media. \emph{Sci. Rep.}, \textbf{4}, 5547.

\bibitem{carmi2007model}
\textsc{Carmi, S., Havlin, S., Kirkpatrick, S., Shavitt, Y.  {\small \&} Shir,
  E.}  (2007) A model of Internet topology using k-shell decomposition.
  \emph{Proc. Natl. Acad. Sci. USA}, \textbf{104}(27), 11150--11154.

\bibitem{zeng2013ranking}
\textsc{Zeng, A.  {\small \&} Zhang, C.-J.}  (2013) Ranking spreaders by
  decomposing complex networks. \emph{Phys. Lett. A}, \textbf{377}(14),
  1031--1035.

\bibitem{tang2015identification}
\textsc{Tang, S., Teng, X., Pei, S., Yan, S.  {\small \&} Zheng, Z.}  (2015)
  Identification of highly susceptible individuals in complex networks.
  \emph{Physica A}, \textbf{432}, 363--372.

\bibitem{malliaros2016locating}
\textsc{Malliaros, F.~D., Rossi, M.-E.~G.  {\small \&} Vazirgiannis, M.}
  (2016) Locating influential nodes in complex networks. \emph{Sci. Rep.},
  \textbf{6}, 19307.

\bibitem{sabidussi1966centrality}
\textsc{Sabidussi, G.}  (1966) The centrality index of a graph.
  \emph{Psychometrika}, \textbf{31}(4), 581--603.

\bibitem{freeman1977set}
\textsc{Freeman, L.~C.}  (1977) A set of measures of centrality based on
  betweenness. \emph{Sociometry}, pp. 35--41.

\bibitem{friedkin1991theoretical}
\textsc{Friedkin, N.~E.}  (1991) Theoretical foundations for centrality
  measures. \emph{Am. J. Sociol.}, \textbf{96}(6), 1478--1504.

\bibitem{dangalchev2006residual}
\textsc{Dangalchev, C.}  (2006) Residual closeness in networks. \emph{Physica
  A}, \textbf{365}(2), 556--564.

\bibitem{brin1998anatomy}
\textsc{Brin, S.  {\small \&} Page, L.}  (1998) The anatomy of a large-scale
  hypertextual web search engine. \emph{Computer Networks and ISDN Systems},
  \textbf{30}(1-7), 107--117.

\bibitem{lu2011leaders}
\textsc{L{\"u}, L., Zhang, Y.-C., Yeung, C.~H.  {\small \&} Zhou, T.}  (2011)
  Leaders in social networks, the delicious case. \emph{PLoS One},
  \textbf{6}(6), e21202.

\bibitem{travenccolo2008accessibility}
\textsc{Traven{\c{c}}olo, B. A.~N.  {\small \&} Costa, L. d.~F.}  (2008)
  Accessibility in complex networks. \emph{Phys. Lett. A}, \textbf{373}(1),
  89--95.

\bibitem{bonacich1972factoring}
\textsc{Bonacich, P.}  (1972) Factoring and weighting approaches to status
  scores and clique identification. \emph{J. Math. Sociol.}, \textbf{2}(1),
  113--120.

\bibitem{radicchi2016leveraging}
\textsc{Radicchi, F.  {\small \&} Castellano, C.}  (2016) Leveraging
  percolation theory to single out influential spreaders in networks.
  \emph{Phys. Rev. E}, \textbf{93}(6), 062314.

\bibitem{martin2014localization}
\textsc{Martin, T., Zhang, X.  {\small \&} Newman, M. E.~J.}  (2014)
  Localization and centrality in networks. \emph{Phys. Rev. E}, \textbf{90}(5),
  052808.

\bibitem{restrepo2006characterizing}
\textsc{Restrepo, J.~G., Ott, E.  {\small \&} Hunt, B.~R.}  (2006)
  Characterizing the dynamical importance of network nodes and links.
  \emph{Phys. Rev. Lett.}, \textbf{97}(9), 094102.

\bibitem{katz1953new}
\textsc{Katz, L.}  (1953) A new status index derived from sociometric analysis.
  \emph{Psychometrika}, \textbf{18}(1), 39--43.

\bibitem{bauer2012identifying}
\textsc{Bauer, F.  {\small \&} Lizier, J.~T.}  (2012) Identifying influential
  spreaders and efficiently estimating infection numbers in epidemic models: A
  walk counting approach. \emph{EPL}, \textbf{99}(6), 68007.

\bibitem{klemm2012measure}
\textsc{Klemm, K., Serrano, M.~{\'A}., Egu{\'\i}luz, V.~M.  {\small \&}
  San~Miguel, M.}  (2012) A measure of individual role in collective dynamics.
  \emph{Sci. Rep.}, \textbf{2}, 292.

\bibitem{lawyer2015understanding}
\textsc{Lawyer, G.}  (2015) Understanding the influence of all nodes in a
  network. \emph{Sci. Rep.}, \textbf{5}, 8665.

\bibitem{gu2017ranking}
\textsc{Gu, J., Lee, S., Saram{\"a}ki, J.  {\small \&} Holme, P.}  (2017)
  Ranking influential spreaders is an ill-defined problem. \emph{EPL},
  \textbf{118}(6), 68002.

\bibitem{borge2012absence}
\textsc{Borge-Holthoefer, J.  {\small \&} Moreno, Y.}  (2012) Absence of
  influential spreaders in rumor dynamics. \emph{Phys. Rev. E}, \textbf{85}(2),
  026116.

\bibitem{chami2017diffusion}
\textsc{Chami, G.~F., Kontoleon, A.~A., Bulte, E., Fenwick, A., Kabatereine,
  N.~B., Tukahebwa, E.~M.  {\small \&} Dunne, D.~W.}  (2017) Diffusion of
  treatment in social networks and mass drug administration. \emph{Nat. Comm.},
  \textbf{8}(1), 1929.

\bibitem{karrer2010message}
\textsc{Karrer, B.  {\small \&} Newman, M. E.~J.}  (2010) Message passing
  approach for general epidemic models. \emph{Phys. Rev. E}, \textbf{82}(1),
  016101.

\bibitem{min2018identifying}
\textsc{Min, B.}  (2018) Identifying an influential spreader from a single seed
  in complex networks via a message-passing approach. \emph{Eur. Phys. J. B},
  \textbf{91}(1), 18.

\bibitem{radicchi2017fundamental}
\textsc{Radicchi, F.  {\small \&} Castellano, C.}  (2017) Fundamental difference between superblockers and
  superspreaders in networks. \emph{Phys. Rev. E}, \textbf{95}(1), 012318.

\bibitem{Jankowski2017balancing}
\textsc{Jankowski, J., Br\'{o}dka, P., Kazienko, P., Szymanski, B. K., Michalski, R. {\small \&} Kajdanowicz, T.} (2017). Balancing speed and coverage by sequential seeding in complex networks. \emph{Sci. Rep.}, \textbf{7}, 891.

\bibitem{Jankowski2018probing}
\textsc{Jankowski, J., Szymanski, B. K., Kazienko, P., Michalski, R. {\small \&} Br\'{o}dka, P.} (2018). Probing limits of information spread with sequential seeding. \emph{Sci. Rep.}, \textbf{8}, 13996.

\bibitem{erkol2019systematic}
\textsc{Erkol, S., Castellano, C. {\small \&} Radicchi, F.}  (2019). Systematic comparison between methods for the detection of influential spreaders in complex networks. \emph{arXiv preprint arXiv:1904.08457}.

\bibitem{rogers2010diffusion}
\textsc{Rogers, E.~M.}  (2010) \emph{Diffusion of innovations}. Simon and
  Schuster, New York.

\bibitem{centola2010spread}
\textsc{Centola, D.}  (2010) The spread of behavior in an online social network
  experiment. \emph{Science}, \textbf{329}(5996), 1194--1197.
  
\bibitem{singh2013threshold}
\textsc{Singh, P., Sreenivasan, S., Szymanski, B. K. {\small \&} Korniss, G.} (2013). Threshold-limited spreading in social networks with multiple initiators. \emph{Sci. Rep.}, \textbf{3}, 2330.

\bibitem{karampourniotis2015impact}
\textsc{Karampourniotis, P. D., Sreenivasan, S., Szymanski, B. K. {\small \&} Korniss, G.} (2015). The impact of heterogeneous thresholds on social contagion with multiple initiators. \emph{PLoS One}, \textbf{10}(11), e0143020.

\bibitem{cohen1997size}
\textsc{Cohen, E.}  (1997) Size-estimation framework with applications to
  transitive closure and reachability. \emph{J. Comput. Syst. Sci.},
  \textbf{55}(3), 441--453.

\bibitem{Karampourniotis2019influence}
\textsc{Karampourniotis, P. D., Szymanski, B. K. {\small \&} Korniss, G.} (2019). Influence Maximization for Fixed Heterogeneous Thresholds. \emph{Sci. Rep.}, \textbf{9}, 5573.

\bibitem{mezard2002random}
\textsc{M{\'e}zard, M.  {\small \&} Zecchina, R.}  (2002) Random
  k-satisfiability problem: From an analytic solution to an efficient
  algorithm. \emph{Phys. Rev. E}, \textbf{66}(5), 056126.

\bibitem{pei2017efficient}
\textsc{Pei, S., Teng, X., Shaman, J., Morone, F.  {\small \&} Makse, H.~A.}
  (2017) Efficient collective influence maximization in cascading processes
  with first-order transitions. \emph{Sci. Rep.}, \textbf{7}, 45240.

\bibitem{janson2008dismantling}
\textsc{Janson, S.  {\small \&} Thomason, A.}  (2008) Dismantling sparse random
  graphs. \emph{Comb. Probab. Comput.}, \textbf{17}(2), 259--264.

\bibitem{sun2018lower}
\textsc{Sun, J., Liu, R., Fan, Z., Xie, J., Ma, X.  {\small \&} Hu, Y.}  (2018)
  Lower bound of network dismantling problem. \emph{Chaos}, \textbf{28}(6),
  063128.

\bibitem{coja2015contagious}
\textsc{Coja-Oghlan, A., Feige, U., Krivelevich, M.  {\small \&} Reichman, D.}
  (2015) Contagious sets in expanders. in \emph{Proceedings of the 26th Annual
  ACM-SIAM Symposium on Discrete Algorithms}, pp. 1953--1987. Society for
  Industrial and Applied Mathematics.

\bibitem{wormald1995differential}
\textsc{Wormald, N.~C.}  (1995) Differential equations for random processes and
  random graphs. \emph{Ann. Appl. Probab.}, \textbf{5}(4), 1217--1235.

\bibitem{wormald1999differential}
\textsc{Wormald, N.~C.} (1999) The differential equation method for random graph
  processes and greedy algorithms. \emph{Lectures on approximation and
  randomized algorithms}, \textbf{73}, 155.

\bibitem{wandelt2018comparative}
\textsc{Wandelt, S., Sun, X., Feng, D., Zanin, M.  {\small \&} Havlin, S.}
  (2018) A comparative analysis of approaches to network-dismantling.
  \emph{Sci. Rep.}, \textbf{8}(1), 13513.

\bibitem{holme2012temporal}
\textsc{Holme, P.  {\small \&} Saram{\"a}ki, J.}  (2012) Temporal networks.
  \emph{Phys. Rep.}, \textbf{519}(3), 97--125.

\bibitem{boccaletti2014structure}
\textsc{Boccaletti, S., Bianconi, G., Criado, R., Del~Genio, C.~I.,
  G{\'o}mez-Gardenes, J., Romance, M., Sendina-Nadal, I., Wang, Z.  {\small \&}
  Zanin, M.}  (2014) The structure and dynamics of multilayer networks.
  \emph{Phys. Rep.}, \textbf{544}(1), 1--122.

\bibitem{kivela2014multilayer}
\textsc{Kivel{\"a}, M., Arenas, A., Barthelemy, M., Gleeson, J.~P., Moreno, Y.
  {\small \&} Porter, M.~A.}  (2014) Multilayer networks. \emph{J Complex
  Networks}, \textbf{2}(3), 203--271.

\bibitem{orsini2015quantifying}
\textsc{Orsini, C., Dankulov, M.~M., Colomer-de Sim{\'o}n, P., Jamakovic, A.,
  Mahadevan, P., Vahdat, A., Bassler, K.~E., Toroczkai, Z., Bogun{\'a}, M.,
  Caldarelli, G.  et~al.}  (2015) Quantifying randomness in real networks.
  \emph{Nat. Comm.}, \textbf{6}, 8627.

\bibitem{erkol2018identifying}
\textsc{Erkol, {\c{S}}., Faqeeh, A.  {\small \&} Radicchi, F.}  (2018)
  Influence maximization in noisy networks. \emph{Europhys. Lett.},
  \textbf{123}(5), 58007.

\bibitem{iribarren2009impact}
\textsc{Iribarren, J.~L.  {\small \&} Moro, E.}  (2009) Impact of human
  activity patterns on the dynamics of information diffusion. \emph{Phys. Rev.
  Lett.}, \textbf{103}(3), 038702.

\bibitem{gallos2012people}
\textsc{Gallos, L.~K., Rybski, D., Liljeros, F., Havlin, S.  {\small \&} Makse,
  H.~A.}  (2012) How people interact in evolving online affiliation networks.
  \emph{Phys. Rev. X}, \textbf{2}(3), 031014.

\bibitem{muchnik2013origins}
\textsc{Muchnik, L., Pei, S., Parra, L.~C., Reis, S.~D., Andrade~Jr, J.~S.,
  Havlin, S.  {\small \&} Makse, H.~A.}  (2013) Origins of power-law degree
  distribution in the heterogeneity of human activity in social networks.
  \emph{Sci. Rep.}, \textbf{3}, 1783.

\bibitem{min2015finding}
\textsc{Min, B., Liljeros, F.  {\small \&} Makse, H.~A.}  (2015) Finding
  influential spreaders from human activity beyond network location. \emph{PLoS
  One}, \textbf{10}(8), e0136831.

\bibitem{aral2012identifying}
\textsc{Aral, S.  {\small \&} Walker, D.}  (2012) Identifying influential and susceptible members of
  social networks. \emph{Science}, \textbf{337}, 337--341.

\bibitem{aral2011identifying}
\textsc{Aral, S.  {\small \&} Walker, D.}  (2011) Identifying social influence
  in networks using randomized experiments. \emph{IEEE Intell. Syst.},
  \textbf{26}(5), 91--96.

\bibitem{aral2018social}
\textsc{Aral, S.  {\small \&} Dhillon, P.~S.}  (2018) Social influence
  maximization under empirical influence models. \emph{Nat. Hum. Behav.},
  \textbf{2}, 375--382.

\bibitem{pei2015exploring}
\textsc{Pei, S., Muchnik, L., Tang, S., Zheng, Z.  {\small \&} Makse, H.~A.}
  (2015) Exploring the complex pattern of information spreading in online blog
  communities. \emph{PLoS One}, \textbf{10}(5), e0126894.

\bibitem{teng2014individual}  
\textsc{Teng, X., Yan, S., Tang, S., Pei, S., Li, W. {\small \&} Zheng, Z.} (2014) Individual 
behavior and social wealth in the spatial public goods game. 
\emph{Physica A}, \textbf{402}, 141--149.

\bibitem{aral2009distinguishing}
\textsc{Aral, S., Muchnik, L.  {\small \&} Sundararajan, A.}  (2009)
  Distinguishing influence-based contagion from homophily-driven diffusion in
  dynamic networks. \emph{Proc. Natl. Acad. Sci. USA}, \textbf{106}(51),
  21544--21549.

\bibitem{centola2011experimental}
\textsc{Centola, D.}  (2011) An experimental study of homophily in the adoption
  of health behavior. \emph{Science}, \textbf{334}(6060), 1269--1272.

\bibitem{aral2013engineering}
\textsc{Aral, S., Muchnik, L.  {\small \&} Sundararajan, A.}  (2013) Engineering social contagions: Optimal network
  seeding in the presence of homophily. \emph{Network Science}, \textbf{1}(2),
  125--153.

\bibitem{centola2007complex}
\textsc{Centola, D.  {\small \&} Macy, M.}  (2007) Complex contagions and the
  weakness of long ties. \emph{Am. J. Sociol.}, \textbf{113}(3), 702--734.

\bibitem{muchnik2013social}
\textsc{Muchnik, L., Aral, S.  {\small \&} Taylor, S.~J.}  (2013) Social
  influence bias: A randomized experiment. \emph{Science}, \textbf{341}(6146),
  647--651.


\end{thebibliography}
%
% once the .bbl file has been generated then place the text in your article.

% To get the numbered reference style the author should use [numbib]
%as an option in the document class.  For example: \documentclass[numbib]{imaiai}

\ifx\undefined\textsc
\newcommand{\textsc}[1]{{\sc #1}}
\newcommand{\emph}[1]{{\em #1\/}}
\let\tmpsmall\small
\renewcommand{\small}{\tmpsmall\sc}
\fi

\end{document}